\documentclass[prapplied, preprint, amsmath, amssymb, aps,superscriptaddress,twocolumn,10pt]{revtex4-2}

\usepackage{graphicx}
\usepackage{dcolumn}
\usepackage{bm}
\usepackage{hyperref}
\usepackage{amsmath}
\usepackage{xcolor}
\usepackage{comment}

\usepackage{hyperref}
\usepackage[capitalize]{cleveref}

\hypersetup{
    colorlinks=true,
    linkcolor=blue,
    linktoc=all,
    filecolor=blue,      
    urlcolor=blue,
    citecolor=blue,
}
\begin{document}
\raggedbottom

\title{A Scalable Stacked-Electrode Printed Circuit Board Radio-Frequency Quadrupole (PCB-RFQ) for Precision Experiments}


\author{Tayemar K. Fowler-Davis}
\affiliation{School of Physics and Astronomy, The University of Edinburgh, Edinburgh, EH9 3FD, United Kingdom}

\author{Moritz Pascal Reiter}
\email[Corresponding author.\newline]{mreiter@ed.ac.uk}
\affiliation{School of Physics and Astronomy, The University of Edinburgh, Edinburgh, EH9 3FD, United Kingdom}

\author{Nawaf Altasan}
\altaffiliation[Present address: ]{Department of Physics, King Saud University, Riyadh, Saudi Arabia}
\affiliation{School of Physics and Astronomy, The University of Edinburgh, Edinburgh, EH9 3FD, United Kingdom}


\author{Samuel Ayet San Andr\'es}
\affiliation{Instituto de F\'isica Corpuscular, CSIC, Universidad de Valencia, Valencia, 46980, Spain}

\author{Peter Black}
\affiliation{School of Physics and Astronomy, The University of Edinburgh, Edinburgh, EH9 3FD, United Kingdom}

\author{Jason Breyiannis}
\affiliation{School of Physics and Astronomy, The University of Edinburgh, Edinburgh, EH9 3FD, United Kingdom}

\author{Callum L. Brown}
\affiliation{School of Physics and Astronomy, The University of Edinburgh, Edinburgh, EH9 3FD, United Kingdom}

\author{Peter Dasiukevich}
\affiliation{School of Physics and Astronomy, The University of Edinburgh, Edinburgh, EH9 3FD, United Kingdom}

\author{Adam Zaki Davies}
\affiliation{School of Physics and Astronomy, The University of Edinburgh, Edinburgh, EH9 3FD, United Kingdom}

\author{Timo Dickel}
\affiliation{GSI Helmholtzzentrum f{\"u}r Schwerionenforschung GmbH, Darmstadt, D-64291, Germany}
\affiliation{II.~Physikalisches Institut, Justus-Liebig-Universit{\"a}t Gie{\ss}en, Gie{\ss}en, D-35392, Germany}


\author{Oscar Hall}
\affiliation{School of Physics and Astronomy, The University of Edinburgh, Edinburgh, EH9 3FD, United Kingdom}

\author{Alexandru Hau}
\affiliation{School of Physics and Astronomy, The University of Edinburgh, Edinburgh, EH9 3FD, United Kingdom}

\author{Jamie C. Jones}
\affiliation{School of Physics and Astronomy, The University of Edinburgh, Edinburgh, EH9 3FD, United Kingdom}

\author{Jan Kocka}
\affiliation{School of Physics and Astronomy, The University of Edinburgh, Edinburgh, EH9 3FD, United Kingdom}

\author{Gabriella Kripk\'o-Koncz}
\affiliation{School of Physics and Astronomy, The University of Edinburgh, Edinburgh, EH9 3FD, United Kingdom}
\affiliation{II.~Physikalisches Institut, Justus-Liebig-Universit{\"a}t Gie{\ss}en, Gie{\ss}en, D-35392, Germany}

\author{Konrad Linkowski}
\affiliation{School of Physics and Astronomy, The University of Edinburgh, Edinburgh, EH9 3FD, United Kingdom}

\author{Adam J. McCarter}
\altaffiliation[Present address: ]{Van Swinderen Institute for Particle Physics and Gravity, University of Groningen, Groningen, 9747 AA, The Netherlands}
\affiliation{School of Physics and Astronomy, The University of Edinburgh, Edinburgh, EH9 3FD, United Kingdom}

\author{Sophia Scrimshaw}
\affiliation{School of Physics and Astronomy, The University of Edinburgh, Edinburgh, EH9 3FD, United Kingdom}

\author{Joe Simon}
\affiliation{School of Physics and Astronomy, The University of Edinburgh, Edinburgh, EH9 3FD, United Kingdom}

\author{Jack Lee Smith}
\affiliation{School of Physics and Astronomy, The University of Edinburgh, Edinburgh, EH9 3FD, United Kingdom}

\author{Wolfgang R. Pla\ss}
\affiliation{II.~Physikalisches Institut, Justus-Liebig-Universit{\"a}t Gie{\ss}en, Gie{\ss}en, D-35392, Germany}
\affiliation{GSI Helmholtzzentrum f{\"u}r Schwerionenforschung GmbH, Darmstadt, D-64291, Germany}

\author{Gemma Robertson}
\affiliation{School of Physics and Astronomy, The University of Edinburgh, Edinburgh, EH9 3FD, United Kingdom}

\author{Jiajun Yu}
\thanks{Deceased}
\affiliation{GSI Helmholtzzentrum f{\"u}r Schwerionenforschung GmbH, Darmstadt, D-64291, Germany}

\author{Alexandra Zadvornaya}
\affiliation{School of Physics and Astronomy, The University of Edinburgh, Edinburgh, EH9 3FD, United Kingdom}
\affiliation{II.~Physikalisches Institut, Justus-Liebig-Universit{\"a}t Gie{\ss}en, Gie{\ss}en, D-35392, Germany}

\date{\today}


\begin{abstract}
Linear radio-frequency quadrupole (RFQ) traps are crucial for ion and phase-space manipulation across diverse physics platforms, including quantum information processing, precision atomic spectroscopy, and high-resolution mass or laser spectrometry. We present the design, electrostatic field optimization, and performance characterization of a scalable, multi-layer printed circuit board (PCB) linear RFQ trap. By utilizing a PCB-based "stacked-electrode" geometry to generate high-quality quadrupolar fields, this architecture suppresses higher-order multipole field components by up to an order of magnitude compared to planar "flat surface-electrode" designs and allows for reclaiming up to $ 60\% $ of the radial pseudopotential-well depth of an ideal hyperbolic quadrupole. Following a thorough optimisation and characterization using both simulations and experiments, we demonstrate its suitability as a cooler buncher, showing rapid helium buffer-gas cooling with time constants between $ 34 \pm 3\ \mu\text{s} $ and $ 412 \pm 23\ \mu\text{s} $, and achieving a highly compressed longitudinal phase-space emittance of only $ 58 \pm 4\ \text{eV}\cdot\text{ns} $. The low beam emittance provides flexible control of the extracted bunch properties: weak extraction fields yield energy spreads down to $ 2.5 \pm 0.4\ \text{eV} $, whereas strong fields produce ultra-narrow temporal widths down to $ 2.6 \pm 0.3\ \text{ns} $. The results establish our PCB-RFQ platform as a versatile, scalable, and cost-effective alternative to traditionally machined rod-based RFQ assemblies for advanced ion-trapping, beam-preparation, or quantum applications.
\end{abstract}

\keywords{Radio-frequency quadrupole, RFQ buncher, Ion cooling, Printed Circuit Boards, Longitudinal emittance, Mass Spectrometry, Collinear Laser Spectroscopy}

\maketitle

\section{Introduction}
\label{Intro}

Trapped-ion applications, ranging from quantum information processing \cite{PhysRevLett.96.253003, hughes2011microfabricated, Xu:2025aa} and optical frequency standards \cite{prestage1989new} to high-resolution mass spectrometry \cite{BLAUM20061} or laser spectroscopy \cite{Neugart_2017}, rely on specialized high-performance ion traps. In many of these applications, continuous ion beams must be thermalized and compressed into highly localized ion assemblies or bunches. To produce ion bunches with a narrow phase-space, the use of small-sized ion traps has been shown to be highly beneficial. 
However, conventional beam preparation relies on macroscopic radio-frequency quadrupole (RFQ) traps machined from solid metals \cite{DEHMELT196853}. Shrinking these macroscopic traps imposes stringent mechanical constraints, as machining tolerances and complex electrode alignment limit the physical scaling of rod-based assemblies.

\begin{figure}[tb!]
\centering
\includegraphics[width=\columnwidth]{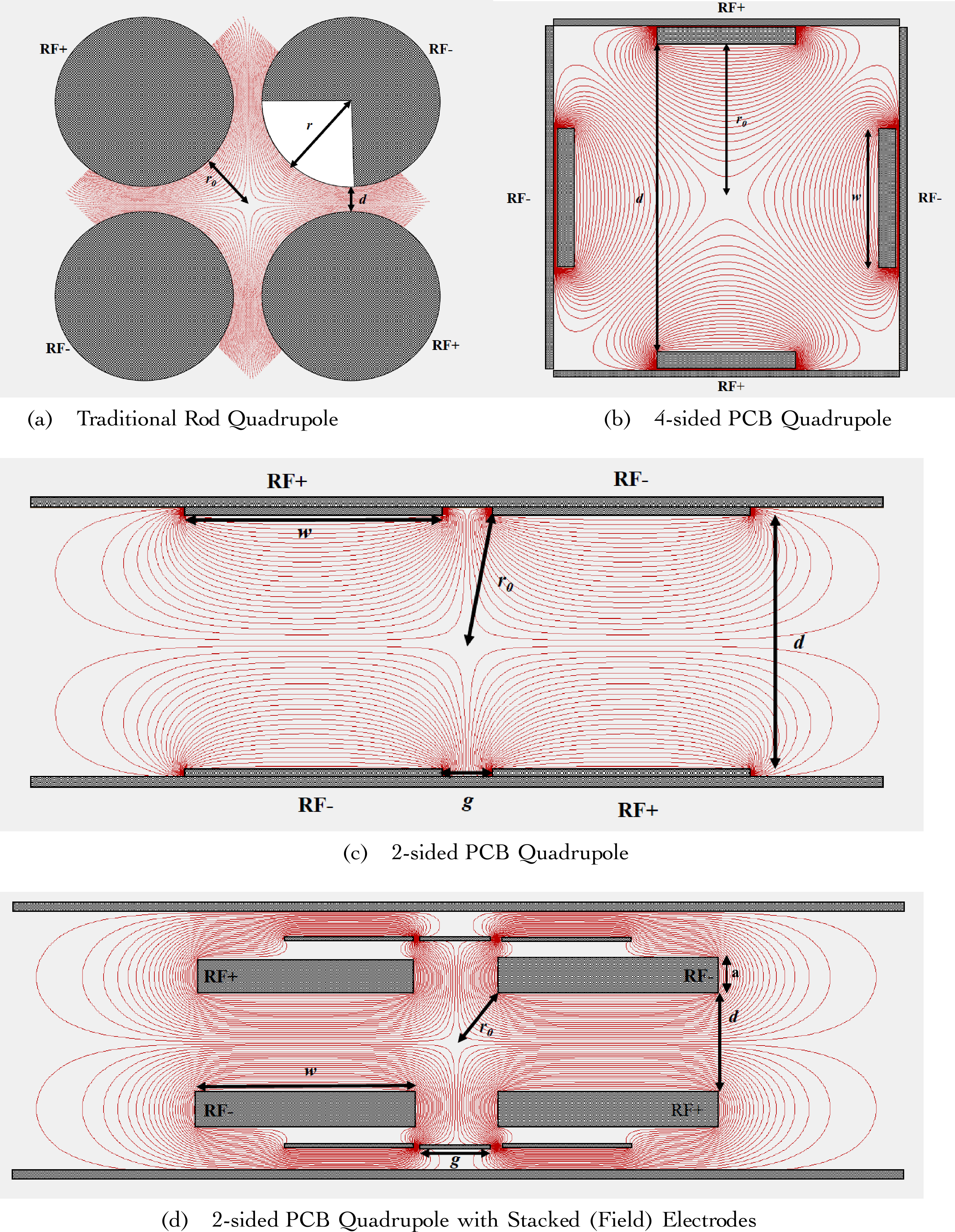}
\caption{Comparison of the field lines for different radio-frequency quadrupole geometries. a) traditional cylindrical rod quadrupole, b) four-sided PCB quadrupole, c) ``flat'' two-sided PCB quadrupole and d) ``stacked'' two-sided PCB quadrupole including additional field electrodes.}\label{fig:FieldLines}
\end{figure}

Advancements in printed circuit board (PCB) technology \cite{coombs2016printed} have lead to an alternative route for ion-optics manufacturing \cite{PhysRevA.75.015401, Jiang_PCB_massfilter, Schlottmann2019}. Multi-layer PCBs leverage standard industrial lithography to deliver high-precision etching (with routine tolerances of $ 10\ \mu\text{m} $ to $ 25\ \mu\text{m} $), and provide inherent alignment precision on the micrometer scale \cite{Jiang_PCB_massfilter, GAMAGE2020116344}. Furthermore, narrow photolithographic traces allow for the fine segmentation of the PCB assembly. Building on these benefits, various planar ion trap designs \cite{SONG2006631, PhysRevA.72.013405} have been developed using parallel-plane or four-sided geometries \cite{ZHANG2017297, CHENG2024100364, atoms11110139, 10.1119/5.0243389} (see Fig.~\ref{fig:FieldLines}). Planar configurations have successfully served as micro-fabricated platforms for quantum computers \cite{PhysRevLett.96.253003, hughes2011microfabricated, Stick:2006aa, Xu:2025aa}, segmented ion guides \cite{Zhang:2015aa, ALLERS201932, Schlottmann2019}, mass analysers \cite{Jiang_PCB_massfilter, Li:2009aa, ZHANG2017297, Decker:2019aa}, and ion bunchers \cite{ITO2013544, Cooper2019, atoms11110139}.

However, transitioning from bulk rods to flat planar electrodes fundamentally alters the electrostatic boundary conditions. Consequently, many planar designs are plagued by higher-order field distortions that induce non-linear resonance effects and transmission losses \cite{l1cn-28kv, Busch:1961aa, FRANZEN199415}. While some implementations achieve high quality quadrupole fields by adding correction electrodes \cite{Jiang_PCB_massfilter, ZHANG2017297, Decker:2019aa} or layers \cite{Bautista-Salvador_2019}, many planar geometries suffer from substantially reduced radial pseudopotential trap depth \cite{PhysRevLett.96.253003}. Because single-board or flat-sandwich designs \cite{Tian:2018aa, GAMAGE2020116344} often yield significantly weaker radial ion confinement and beam acceptance \cite{Stick:2006aa, Decker:2019aa, Xu:2025aa}, they are often restricted to manipulating ions that are already cold (pre-cooled). Standard planar PCBs lack the radial confinement necessary to prevent transverse ion loss when applying steep axial trapping potentials and collisional buffer-gas cooling 
\cite{Moriwaki_1992, PhysRevA.75.015401, 10.1063/1.3665647}.

In this work, we present a novel ''stacked-electrode'' printed circuit board architecture that addresses the inherent limitations of planar ''flat-electrode'' designs. By introducing additional, vertically extended PCB field-electrodes into a standard two-board sandwich trap, we improve the electrostatic boundary conditions without sacrificing mechanical simplicity. This 'stacked'' geometry suppresses unmitigated higher-order multipoles by up to an order of magnitude and significantly strengthens the radial pseudopotential depth compared to a simple ''flat-electrode'' design, ultimately reclaiming up to $ 60\% $ of the trapping well of an ideal hyperbolic quadrupole.

To demonstrate the robustness and field quality of this architecture, we deploy it as a linear cooler-buncher. Fully integrated with on-board resistor networks, capacitive coupling, and an internal high-vacuum heating layer, our stacked device successfully bridges the gap between deep radial confinement and PCB scalability. The device achieves highly compressed longitudinal phase-space emittances, permitting flexible and dynamic control over the extracted bunch properties. By seamlessly delivering either narrow temporal widths or low energy spreads, our results establish this easy-to-manufacture PCB architecture as a highly versatile and powerful replacement for traditionally machined rod-based assemblies across advanced ion-trapping and beam-preparation applications.

\section{Principles of RFQ Cooler-Bunchers}
\label{RFQs}

The fundamental necessity for beam-preparation devices with intrinsically low longitudinal emittance is particularly pronounced at radioactive ion beam (RIB) facilities. Here, low-intensity beams of short-lived exotic isotopes are initially produced with inherently large phase-space volumes and high internal temperatures, making high-performance trapping and bunching platforms routine, indispensable instruments at RIB facilities \cite{PhysRevLett.88.094801, isoltrap08, Isolde_laser09, PODADERAALISEDA2004647, BRUNNER201232, risp_2017, SCHWARZ2016131, GERBAUX2023167631, HAETTNER2018138, JARIES2025170273, BARQUEST201718, BARQUEST2016207, VALVERDE2020330}. Because beam emittance acts as an incompressible phase-space volume during ideal (Hamiltonian) beam transport, a cooler-buncher with a highly compressed longitudinal emittance provides critical operational flexibility. It can dynamically tune its pulsed extraction fields to deliver either exceptionally narrow temporal widths, e.g. maximizing the mass-resolving power in Multiple-Reflection Time-Of-Flight Mass Spectrometers (MR-TOF-MS) \cite{PLA2013134, WOLF201282,DICKEL2015172,ROSENBUSCH2023167824}, or ultra-low energy spreads, e.g. minimizing Doppler or resonance broadening in collinear fast-beam laser spectroscopy \cite{CAMPBELL2016127,YANG2023104005} or Penning traps \cite{102711_094939}.
The drive for ever-lower beam emittances has motivated the development of compact ion bunchers \cite{HAETTNER2018138,VIRTANEN2025170186,gp2b-krwb, ITO2013544}, as well as increasingly complex beam-preparation techniques, including cryogenic buffer-gas cooling \cite{LECHNER2024169471} and in-trap laser Doppler cooling \cite{Schneider2014PRA,PhysRevResearch.4.033229}. While effective, these approaches introduce severe operational complexity and infrastructure requirements, limiting their applicability and scalability.

The fundamental operation of a linear RFQ cooler-buncher relies on the combination of three distinct mechanisms: a radio-frequency (RF) field for radial confinement, an axial direct-current (DC) field for longitudinal guiding and trapping \cite{DEHMELT196853}, and a neutral buffer gas for dissipative cooling \cite{DEHMELT196853}. 

Ions injected into the device undergo continuous collisions with the buffer gas, systematically losing kinetic energy \cite{DEHMELT196853,Major:1968zz,Moriwaki_1992}. To prevent the ions from scattering outwards and being lost to the electrodes, the RF field creates a strong radial confining force, focusing the thermalized ion cloud along the central axis of the trap. 
Assuming an ideal RFQ geometry comprising four infinite hyperbolic rods, the transverse electric potential $ \Phi(x,y,t) $ can be described analytically in Cartesian coordinates as
\begin{equation}
\Phi(x,y,t) = \Phi_0(t)\,\frac{x^2 - y^2}{r_0^2},
\end{equation}
where $ \Phi_0(t) $ is the applied time-dependent potential (comprising a DC offset $ U $ and a zero-to-peak RF amplitude $ V_{RF} $ oscillating at an angular frequency $ \omega_{\mathrm{RF}} $), and $ r_0 $ is the characteristic inscribed field radius from the central axis to the electrode surfaces.  

In practical applications, however, deviations from infinite hyperbolic electrode shapes and misalignments distort this ideal field and lead to higher-order multipole components \cite{Busch:1961aa,FRANZEN199415,GAMAGE2020116344,Tian:2018aa}. Their fields are best quantified by expressing the continuous 2D transverse potential as a multipole expansion in polar coordinates `$ (r, \theta) $` with coefficients $C_{2n}$:
\begin{equation}
\Phi(r,\theta,t) = \Phi_0(t) \sum_{n=1}^{\infty} C_{2n} \left(\frac{r}{r_0}\right)^{2n} \cos(2n\theta). \label{eq:multipole}
\end{equation}
For a pure harmonic quadrupole, only the $ n=1 $ ($ C_2 $) coefficient is non-zero. In traditional RFQ cooler-buncher assemblies, the ideal hyperbolic surfaces are approximated using circular cylindrical rods with radius $r_\text{rod}$. In these assemblies, odd-$n$ multipole coefficients are suppressed due to inherent symmetries, whereas the first major perturbing term, the dodecapole ($ C_6 $) term, is minimized by choosing an aspect ratio of $ r_\text{rod}/r_0 \approx 1.1451 $ \cite{DEHMELT196853}. In high-performance quadrupole mass filters, aspect ratios in the range of $ r_\text{rod}/r_0 \approx 1.125 $ to $ 1.130 $ are commonly chosen, which introduces a controlled, small, positive, $ C_6 $ term to destructively interfere with higher-order terms, such as the icosapole ($ C_{10} $) coefficient \cite{https://doi.org/10.1002/rcm.735}.

However, due to the inherent mismatch between rectangular and planar traces and hyperbolic surfaces, PCB-based designs introduce significantly larger higher-order multipole terms \cite{Tian:2018aa}. In particular, near the electrode surface of flat PCB-RFQ designs, the ideal quadratic potential is distorted by interference from competing higher-order spatial harmonics. 
If left unmitigated, these terms perturb the trapping potential and can generate resonances within the standard Mathieu stability diagram, ultimately inducing beam losses during ion transport and trapping.

Beyond the geometric shape of the field, the capacity of the RFQ to radially confine the ion cloud is given by its time-averaged effective potential, known as the pseudopotential $\Psi(r,\theta)$. The oscillating electric field with a zero-to-peak amplitude $E_{\mathrm{RF}}(r,\theta)$ causes an ion with mass $m$ and electric charge $e$ to experience a time-averaged restoring force. This force drives the ion toward the symmetry axis ($r=0$) of the RFQ, where the RF pseudopotential is minimal as given by:
\begin{equation}
\Psi(r,\theta) = \frac{e\,E_{\mathrm{RF}}^{2}(r,\theta)}{4\, m \, \omega_{\mathrm{RF}}^{2}}.
\end{equation}

In linear RFQs, this pseudopotential establishes an effective radial trapping well, in which the ion undergoes a slow secular oscillation. For an ideal quadrupole field, the pseudopotential $\Psi(r)$ and its depth $\Psi_{\text{depth}}$, which defines the maximum effective trapping energy available in the RF well, can be calculated as:
\begin{equation}
    \Psi(r) =
\frac{e\,V_{\mathrm{RF}}^{2}}{m \,\omega_{\mathrm{RF}}^{2}\, r_0^{4}}\, r^{2}\textrm{~and~}\Psi_{\text{depth}}
= \Psi(r_0)
= \frac{e\,V_{\mathrm{RF}}^{2}}{m \,\omega_{\mathrm{RF}}^{2}\, r_0^{2}}.
\end{equation}

Due to the weakened and often asymmetric electric field, planar ion traps produce shallow confining potentials \cite{Stick:2006aa,Decker:2019aa,Xu:2025aa}, which makes them highly susceptible to collisional ion losses and often renders them unsuitable for ion cooling and preparation. To successfully replace a traditionally machined RFQ cooler-buncher with a PCB-based design, one must simultaneously minimize higher-order multipole coefficients to ensure stable ion trajectories, and maximize the pseudopotential strength to ensure a sufficiently deep radial trapping well.

Once the ions are radially confined and longitudinally accumulated in a local potential minimum, a steep homogeneous axial extraction field can be applied to eject the thermalized ions as a dense ion bunch.
For an ion cloud with an internal spatial extent along the beam axis, $ \Delta z^{\text{FWHM}} $, the resulting width of the kinetic energy distribution upon extraction, $ \Delta K ^{\text{FWHM}}$, scales linearly with the applied extraction field strength $ E_\mathrm{ext} $:
\begin{equation}
\Delta K^{\text{FWHM}} = e\,E_\mathrm{ext}\,\Delta z^{\text{FWHM}}.\label{eq:DK}
\end{equation}
Concurrently, the minimum fundamental limit to the temporal width of the extracted ion bunch at the focal plane, $\Delta t^{\text{FWHM}}_{\text{ta}}$, is dictated by the ions' turnaround time \cite{10.1063/1.1712366} and, assuming a thermal ion motion at an equilibrium temperature $ T $, it is given by: 
\begin{equation}
\Delta t^{\text{FWHM}}_{\text{ta}} = \frac{\sqrt{8\ln(2)\,m\,k_\textrm{B} T}}{e\, E_\mathrm{ext}},
\label{eq:DT}
\end{equation}
with $k_\textrm{B}$ being the Boltzmann constant.
Because both the achievable temporal and energy spreads, $\Delta t^{\text{FWHM}}$ and $\Delta K^{\text{FWHM}}$, depend strongly on the magnitude of the extraction field, $ E_\mathrm{ext} $, a trade-off emerges. Minimizing the turnaround time to achieve the narrowest possible temporal spread requires a strong extraction field. However, applying this large DC gradient across the spatial spread of the ion cloud unavoidably inflates the final energy spread. The product of these two complementary observables represents the longitudinal emittance, $ \epsilon_\text{long} $, which remains conserved during and after extraction, representing an intrinsic figure of merit for the extraction performance and the overall design quality of an RFQ cooler-buncher.
For true homogeneous extraction from an RFQ cooler–buncher, the RMS longitudinal emittance can be calculated by: 
\begin{equation}
   \epsilon_\text{long}^\text{RMS}  = \pi \cdot \Delta t^{\text{RMS}} \cdot \Delta K^{\text{RMS}} = \frac{\pi  \cdot \Delta t^{\text{FWHM}} \cdot \Delta K^{\text{FWHM}}}{8\ln(2)}
   . \label{eq:emmittance}
\end{equation}
In practice, inhomogeneous and time-varying fields as well as collisions with buffer and residual gas during extraction will broaden the longitudinal emittance.  

As shown in \cref{eq:DK}, the ultimate energy spread of the extracted bunch is proportional to the initial width of the thermalized ion package along the beam axis, $ \Delta z^{\text{FWHM}} $, prior to extraction. 
This spatial compression requires the application of a narrow, steep axial trapping potential well with a depth of $ V_{\text{axial}} $. 
However, establishing a strong, axially confining potential inherently introduces a radially defocusing electrostatic force throughout the trapping volume. To prevent the ions from leaking out radially, the local radial pseudopotential must match the axial trapping potential ($ \Psi_{\text{depth}} > V_{\text{axial}} $)~\cite{HERFURTH2001254, GINS201924}. 
Thus, compressing an ion cloud to a highly localized volume in $z $-direction requires a similarly compressed trap geometry in $r $-direction. 
Achieving state-of-the-art longitudinal phase-space compression benefits from an overall small trapping geometry and indeed fabricating small-sized ion bunchers has emerged as a prime mechanism to minimize longitudinal beam emittance \cite{HAETTNER2018138,VIRTANEN2025170186}.

\section{Design and Construction of a PCB-RFQ}
Due to the high-precision (lithographic) manufacturing of the PCBs themselves, the dominant mechanical intolerances in PCB based RFQs come from the mechanical support and mounting structures. Here, two-sided PCB quadrupole configurations exhibit an inherent simplicity in comparison to four-sided PCB quadrupole designs, as several critical dimensions are locked via the PCB surface, which reduces potential sources of electrode misalignment and misplacement, see discussion in \cite{GAMAGE2020116344,Tian:2018aa}.

In this study, we initially explored a pure ``flat'' two-sided sandwich-type PCB quadrupole configuration (Fig.~\ref{fig:FieldLines}c). However, the subsequent iterations introduced a ``stacked'' two-sided PCB quadrupole configuration (Fig.~\ref{fig:FieldLines}d). A ``flat'' two-sided PCB quadrupole geometry reduces the available degrees of freedom for the electrode design to just three, as shown in Fig.~\ref{fig:FieldLines}c; the horizontal gap `$g$` between the thin field pads and their width `$w$` on the PCB board itself, and the vertical spacing between the two PCB boards `$d$`. However, ``flat'' electrodes, spanning only one dimension in the tranverse plane, inherently struggle to mimic the full two-dimensional hyperbolic fields, motivating the development of a ``stacked'' electrode design.   
Thus, in addition to the `$g,w,d$` parameter space of the ``flat'' PCB geometry, our ``stacked'' two-sided PCB quadrupole geometry introduces the fourth degree of freedom: the thickness `$a$` of an additional PCB field electrode, see Fig.~\ref{fig:FieldLines}d. 

\subsection{Electromagnetic Simulations and Geometry Optimization}
\label{sec:simulations}
To leverage the simplified manufacturing of PCBs, while maintaining high-quality fields, electric field simulations were performed using the SIMION v8.2 software package \cite{DAHL20003}, where geometries were constructed on a $x-y$ pixel mesh with pixel size of $0.01$~mm (vertical) and $0.02$~mm (horizontal), respectively. The calculated electric field $\Phi(x,y)$ was extracted and transformed into $\Phi(r,\theta)$ coordinates.  
Overall, the simulations explored a wide range of `$g,w,d,a$` parameter combinations, with each combination realised through a dedicated SIMION geometry. First, the effect of changing the pad width `$w$` was investigated. The simulations showed that once the pad width was about a factor of two larger than a given vertical spacing `$d$`, the simulated geometries produced horizontal field lines between the RF+ and RF- pads. Increasing the pad width further did not affect the central quadrupole field lines. For the remaining simulations, a pad width `$w$` of $20$~mm was kept. 
Next, the aspect ratio between the vertical spacing `$d$` and horizontal gap `$g$` was optimized. For each aspect ratio and the corresponding electrode geometries, the RMS deviation between the simulated electric fields $\Phi(r,\theta)$ and that of an ideal quadrupole field $\Phi_{c2}(r,\theta)$ was calculated across $\sim 9000$ points in $r_i$ and $\theta_i$, as 
\begin{equation}
\textrm{RMS}_\Phi = \sqrt{ \frac{1}{n} \sum_{i=1}^{n} \left( \Phi(r_i,\theta_i)-\Phi_{c2}(r_i,\theta_i) \right) ^2}
\end{equation}
In addition, to gain a deeper insight into the characteristics of the different geometries, a multipole expansion fit using \cref{eq:multipole} was performed on the $\Phi(r_i,\theta_i)$ points for each aspect ratio \cite{BARLOW200119}, set to determine higher-order multipole coefficients up to the 24th order. For both procedures, the calculations were limited to the inner $50\%$ field radius corresponding to the respective quadrupole geometry, justified by the fact that in typical ion bunching applications thermalized ions predominantly occupy the inner regions of the buncher with $r < 0.4 \, r_0 $, owing the operation at low-to-moderate Mathieu $q$-values \cite{Dickel:200291}.   

\begin{figure}[tb!]
\centering
\includegraphics[width=1\columnwidth]{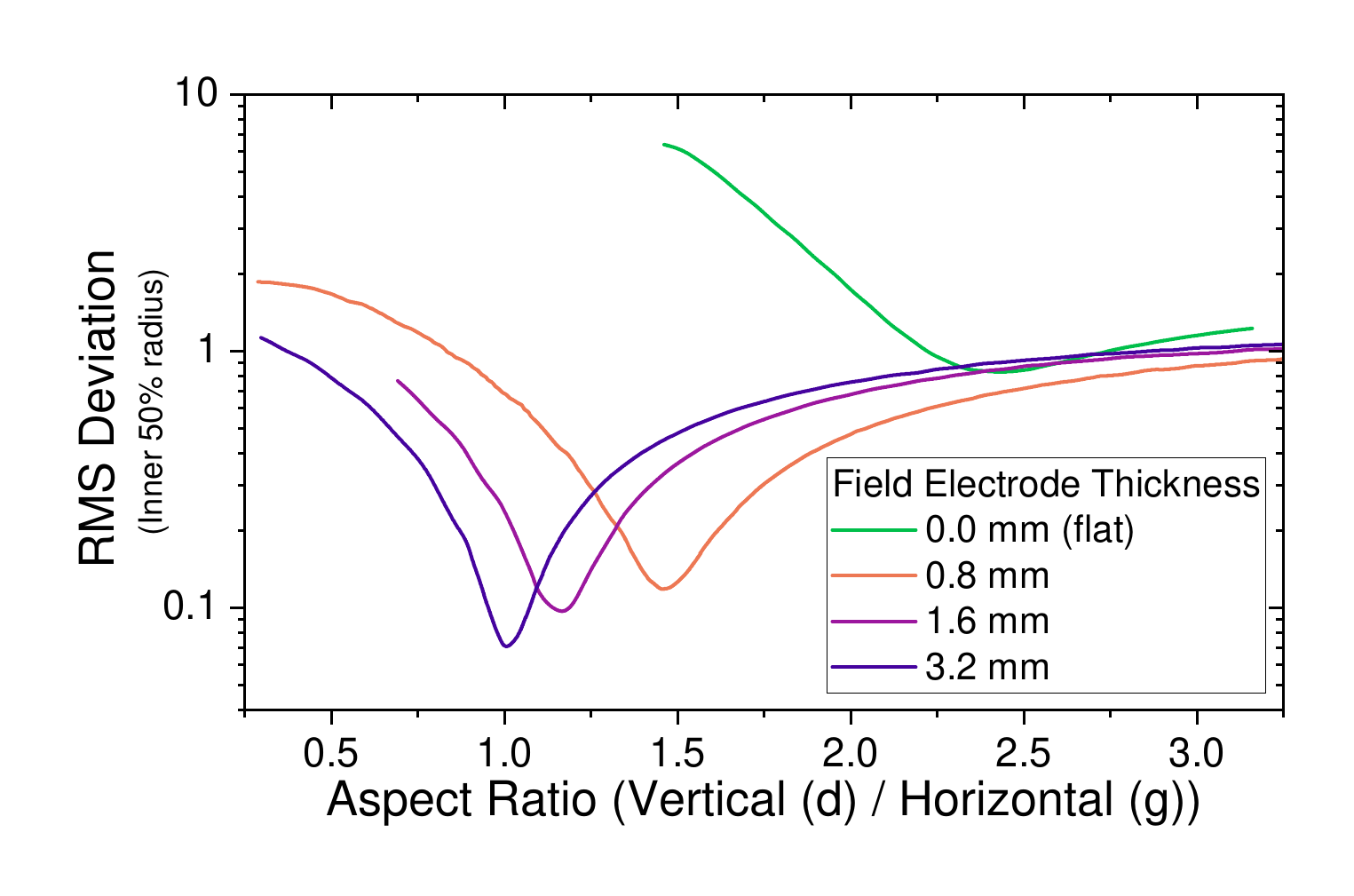}
\caption{RMS deviation between the simulated electric fields and an ideal quadrupole field at different ``flat'' two-sided PCB quadrupole geometries as a function of the aspect ratio (vertical-to-horizontal distance ratio). Increasing the thickness of the field electrodes moves the optimal aspect ratio towards $1:1$, while also reducing the residual non-quadrupole field deviation substantially.}\label{fig:RMS}
\end{figure}

The outcome of the RMS evaluation, adopting a ``flat'' PCB quadrupole design with standard, thin gold-plated copper traces mounted flush with the PCB surface, is shown by the green curve in Fig.~\ref{fig:RMS} as an electrode thickness $a=0$~mm. 
In this ``flat'' geometry, the RMS deviation $\textrm{RMS}_\Phi$ shows a clear minimum at an aspect ratio of $2.4:1$, defined as the ratio of the vertical spacing $d$ to the horizontal gap $g$ between the electrodes and PCB boards. At this aspect ratio, a minimum $\textrm{RMS}_\Phi$ of $\approx 0.82 $ is obtained, which seems sufficient for basic transport of cold ion clouds. 
To improve the symmetry and strength of the produced electric fields, the field electrodes' physical dimensions were artificially increased by introducing an additional PCB layer carrying gold-plated copper traces mounted flush with the added layer, thereby creating our ``stacked'' PCB geometry, as shown in Fig.~\ref{fig:FieldLines}d. \cref{fig:RMS} shows the obtained RMS deviation $\textrm{RMS}_\Phi$ for selected field electrode thicknesses across different aspect ratios. The thicknesses $a$ of the additional field electrodes were selected to match the thickness of commonly available PCB boards with $ 0.8\text{ mm} $, $ 1.6\text{ mm} $ and $ 3.2\text{ mm} $. First, changing to this ``stacked'' topology and increasing the thickness $a$ stepwise shifts the aspect ratio at which $\textrm{RMS}_\Phi$ is minimized towards the intuitively expected ideal aspect ratio of $ 1:1 $ for a geometry assuming infinite square electrodes. Second, for a field electrode thickness of $ a=3.2\text{ mm} $, an aspect ratio of $ 1:1 $ was found to yield a minimum $\textrm{RMS}_\Phi$ deviation of only $\approx  0.068 $, which is more than an order of magnitude smaller than the previous optimum found with the ``flat'' PCB design. 

To assess how severe such remaining $\textrm{RMS}_\Phi$ deviations would be, we performed a simulation of a commonly used RFQ cooler-buncher assembly employing parallel cylindrical rods of radius $r_\text{rod}$, with an aspect ratio $r_\text{rod}/r_0 = 1.145 $. Under otherwise identical parameters, the cylindrical-rod geometry yielded an $\textrm{RMS}_\Phi$ of $ \approx 0.025 $. As such, the $\textrm{RMS}_\Phi$ of our ``stacked'' PCB quadrupole design is only about a factor of $\sim 3$ higher compared to that of commonly used round quadrupole mass filter RFQs.

To better understand the contributions from higher-order field components the results obtained from the multipole expansion fits were examined. Due to their inherent symmetries, both designs suppress odd-$n$ multipole coefficients, but leave meaningful $C_6$, $C_{10}$, $C_{14}$ and $C_{18}$ multipole field components.
For the ``flat'' PCB quadrupole design, the $C_6$, $C_{10}$, $C_{14}$ and $C_{18}$ multipole components remain large, contributing to the field with similar amplitudes and an alternating $ + \ - \ - \ + $ sign pattern even at the minimum RMS aspect ratio of $ d/g \approx 2.4:1 $. 
Because flat boards lack vertical physical boundaries, the electric field lines bulge outward into the horizontal gap between the electrode pads at the PCB boards' surface (\cref{fig:FieldLines}c). To describe these bulging fields and match the potential at the electrode boundaries, the electrostatic fit is forced to include competing, higher-order multipoles. 
 
By transitioning to a 'stacked-electrode' geometry, the additional vertically extended field electrodes reduce the global $\textrm{RMS}_\Phi$ deviation and higher-order multipole contributions significantly. This, more symmetric geometry, was found to reduce the $C_6$, $C_{10}$, $C_{14}$ and $C_{18}$ multipole field components by factors of about 5-to-10 compared to the ``flat'' PCB design. 
However, the inherent mismatch between the now almost rectangular surface electrodes and hyperbolic field lines can not fully be resolved. The expanded vertical faces of the ``stacked'' electrode boards lie closer to the ion-beam axis than the corresponding surfaces of ideal hyperbolic electrodes. Thus, these enlarged vertical faces squeeze the equipotential field lines (\cref{fig:FieldLines}d) and the field solution necessarily includes $C_6$, $C_{10}$, $C_{14}$ and $C_{18}$ multipole components to describe this deformation; effectively bending the potential more sharply to meet the flat electrode corner. Across a narrower range of aspect ratios between $0.88:1 $ and $1.15:1 $, the multipole terms remain stable, which indicates some level of robustness of the ``stacked'' geometry against small real-world aspect ratio variations.    
 
\begin{figure}[tb]
\centering
\includegraphics[width=1\columnwidth]{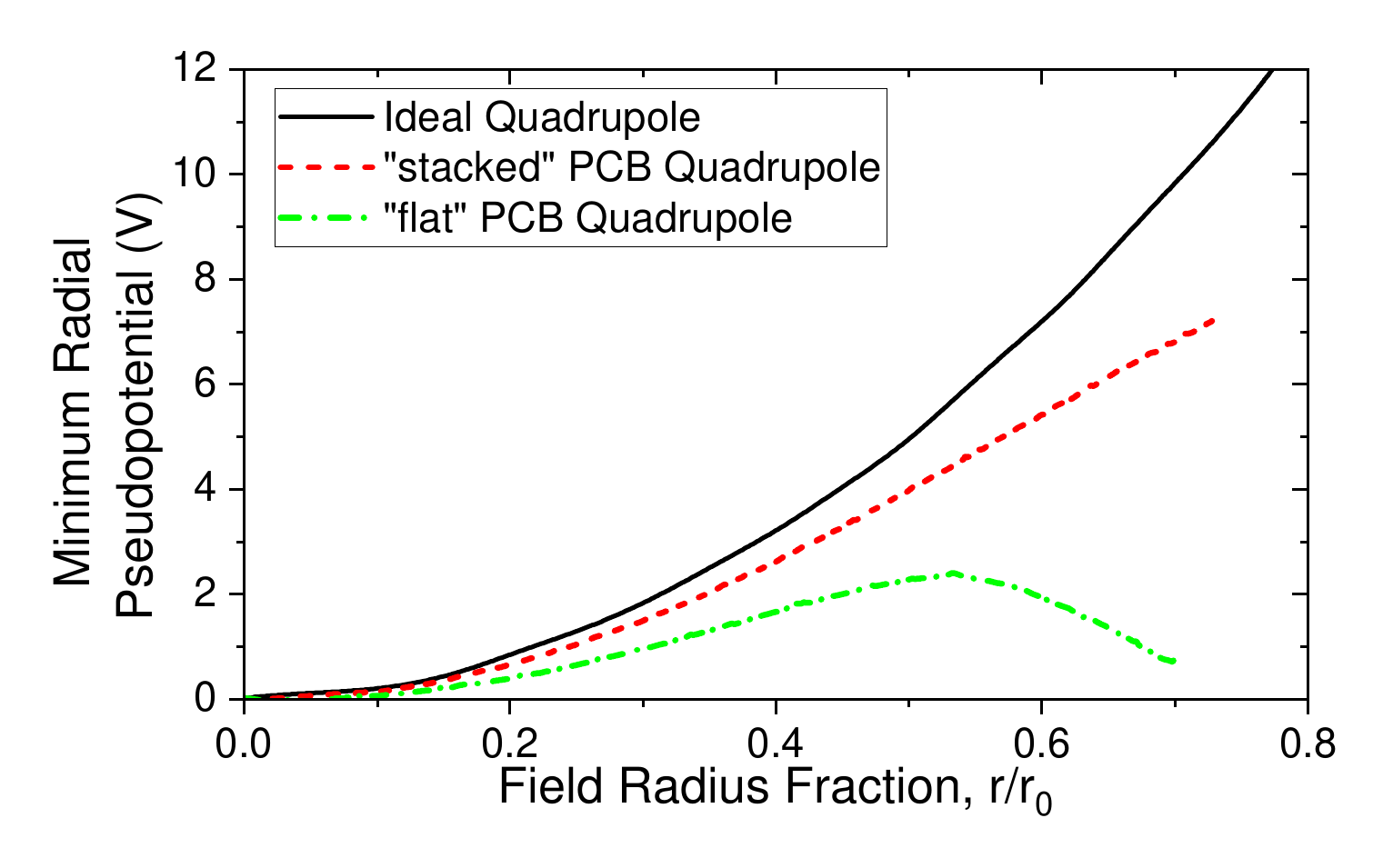}
\caption{Minimum radial confining (time-averaged) pseudopotential (in V) determined based on electric field simulations for different quadrupole ion trap geometries as a function of the field radius fraction $r/r_0$. The improvement of our ``stacked'' PCB-RFQ design over the ``flat'' PCB-based cooler-buncher design is clearly visible.}
\label{fig:pseudo}
\end{figure}

To ensure efficient ion transport, linear RFQs need to achieve a sufficiently strong pseudopotential for radial ion confinement. Based on the electric field simulations described above, the time-averaged transverse potential was assessed across the vertical and horizontal axis. Fig.~\ref{fig:pseudo} shows the minimum time-averaged potential for an ideal quadrupole field, as well as for our optimized ``flat'' and ``stacked'' PCB quadrupoles; all scaled to operate at the same effective operational $q$-value. Due to the planar geometry of the ``flat'' PCB quadrupole, it produces a much less symmetric quadrupole field and hence a reduced confining potential along the horizontal axis. 
Considering the maximum radial trap depth, given by the minimal radial confinement, the ``flat'' PCB only reaches about $18\%$ of the depth of an ideal quadrupole.  
In comparison, the ``stacked'' PCB quadrupole creates a much more uniform and symmetric confining potential, recovering to about $60\% $ of the pseudopotential-well depth of a perfect hyperbolic quadrupole. 
By providing a deeper pseudopotential well, the ``stacked'' design permits the use of steeper axial potential wells, which directly translates into tighter spatial ion compression and a reduced longitudinal beam emittance. 

To establish axial ion transport and local bunching, a DC gradient is commonly applied along a segmented linear RFQ. To determine how many segments are sufficient to achieve a smooth DC gradient, a simulation was performed in which a $ 100\text{-mm} $-long linear RFQ (r$_0=2.88$~mm) was divided into $2$, $5$, $9$, $17$, and $33$ segments and a DC gradient was applied along the segmented RFQ. The results confirm that once the segmentation pitch approaches the field radius $r_0$, non-linearities in the axial transport field vanish, yielding an smooth DC gradient. 

This is where the advantage of PCB-based RFQs compared to traditional segmented metal rods becomes most apparent. The PCB-based RFQ design inherently allows for a straightforward fine segmentation across its length, resulting in smoother DC gradients. This is particularly important for designing linear RFQ ion traps with small field radii $r_0$, which are needed to achieve tightly confined ion clouds and the formation of ion bunches with minimal emittance, while at the same time require a fine axial segmentation.

\subsection{PCB Architecture and PCB-RFQ Setup}
\label{sec:pcb_architecture}
Several prototypes with ``flat'' and ``stacked'' PCB-RFQ electrodes, with field radii $r_0$ between 2-to-5.5~mm were developed before converging to the final ``stacked'' PCB quadrupole design. First design iterations explored the ``flat'' two-sided PCB quadrupole geometry and used surface-mount inductors for RF blocking, requiring separate DC resistor chains for the RF$+$ and RF$-$ rails. Subsequent prototypes transitioned to using resistive RF blocking, which improved the robustness against high-voltage RF, and adopted the stacked-electrode geometry with dedicated field electrode boards to enhance the achievable electric field quality and radial ion confinement in the PCB-based cooler-buncher.

\begin{figure}[tb!]
\centering
\includegraphics[width=\columnwidth]{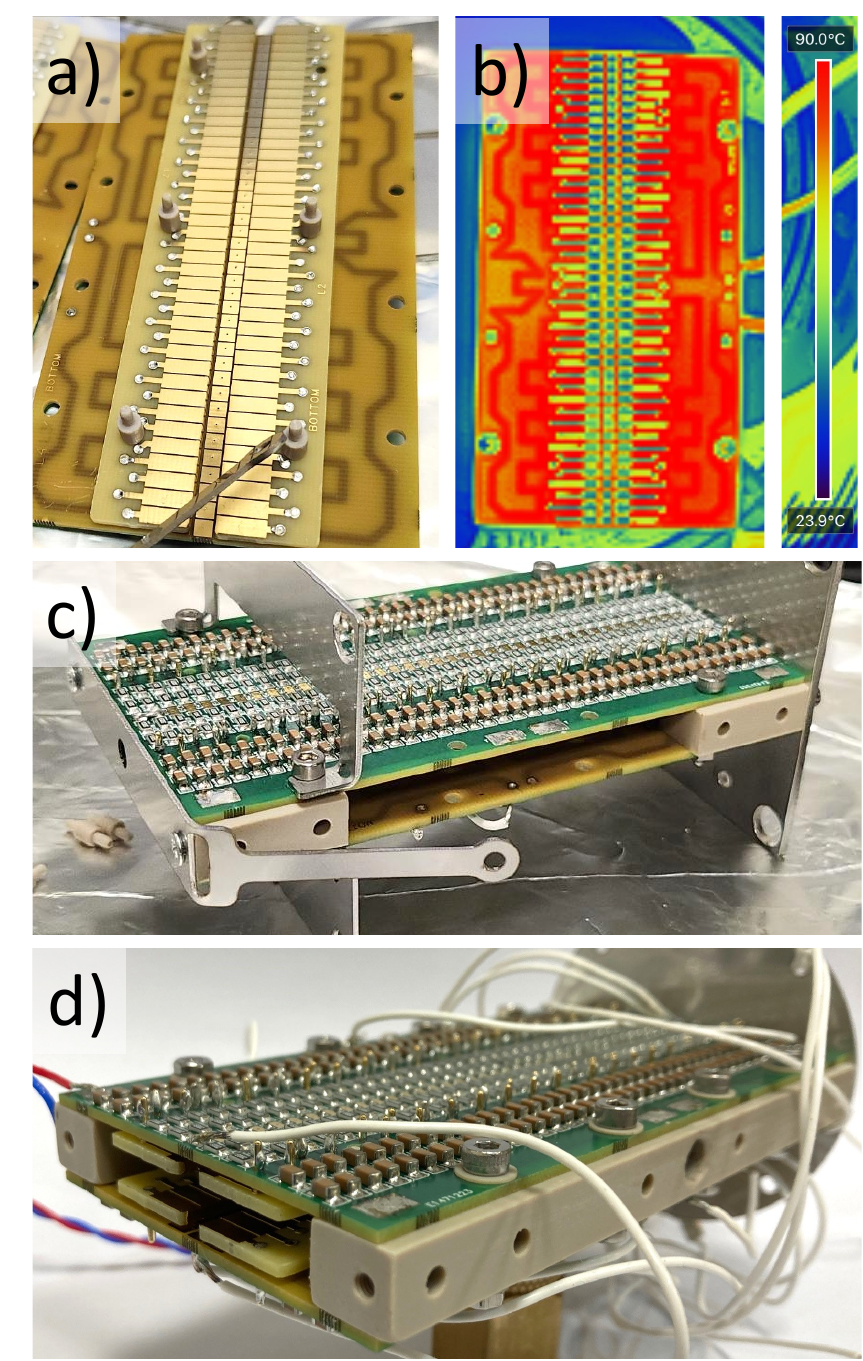}
\caption{Photographs of the designed PCB-RFQ cooler-buncher assemblies showing (a) the exposed field electrodes, (b) a thermal image of the PCB-RFQ applying 15~W heating power in air, (c) a non-encapsulated PCB-RFQ with apertures, and (d) a PCB-RFQ buncher equipped with a side PEEK spacer block for encapsulation and differential pumping.}
\label{fig:buncher}
\end{figure}

The PCB-RFQ cooler-buncher discussed herein was built solely using multi-layer PCB technology. Its design is based on the trap system of the MR-TOF-MS of the FRS Ion Catcher \cite{Dickel:200291}. 
Our device consists of a main segmented PCB that carries $40$ segments (separated by $0.15$~mm gaps) along an active length of $120$~mm. To operate as a two-stage cooler-buncher, an additional one-segment-long PCB-RFQ trap module with a length of $4$~mm was added downstream. The one-segment trap module follows the same electrode geometry as the segments of the main PCB-RFQ, such that the transverse quadrupole field and radial ion confinement provided by the pseudopotential are nominally identical in both sections when operated with the same RF amplitude and frequency. However, the electrical independence of the trap module enables the formation of a compact ion bunching region separated by $1$-mm-thick apertures. The apertures, in return, allow for creating well-defined extraction fields that are as homogeneous as possible across the trapping region. 

Our ``stacked'' PCB-RFQs are based on a six-layer rigid FR-4 architecture (custom-fabricated by Eurocircuits, Belgium), hereafter referred to as the \emph{component board}. This board provides mechanical support and hosts the surface-mount passive network for RF/DC mixing and distribution, populated on the outer face relative to the beam axis. A resistor chain using $ 100\text{ k}\Omega $ resistors generates the axial DC transport gradient by stepping the potentials of the field-electrode segments along the length of the PCB-RFQ, while high-voltage coupling capacitors distribute the RF voltage to each electrode segment and block the DC bias from the resonant RF source. This is achieved by two $ 2200\text{ pF} $ capacitors coupled in series to the RF$+$ and RF$-$ rails on each side of the \emph{component board}. Vice versa, the DC supplies are shielded from the RF voltage using two $ 2.2\text{ M}\Omega $ resistors coupled in series. The network is implemented symmetrically for each segment and for the two RF phases around the central resistor chain. To minimise the risk of discharge, the trace and pad clearances, and the via placement were designed to allow for high-voltage operation within a vacuum environment. The PCB-RFQ design was tested so far up to DC floating voltages of $1.5$~kV and peak-to-peak RF amplitudes of $ 800\text{ V}_{pp} $. 

In order to improve the vacuum suitability and outgassing characteristics of the FR-4 base \cite{Conseil-Gudla:2017aa}, no solder mask was applied to the inner facing side of the \emph{component board} and one of its inner layers was equipped with an internal heating trace (the trace is visible in Fig.~\ref{fig:buncher}a) and a PT100 temperature sensor was placed for in-situ temperature monitoring. Fig.~\ref{fig:buncher}b shows a thermal image of our PCB RFQ applying 15~W of power (5~A at 3~V) in air. The whole board can be heated homogeneously to $< 80$~C$^{\circ}$, fostering improved outgassing of absorbed water. In the future, our PCB-RFQ cooler-buncher will be built using ceramic-filled PTFE boards (such as Rogers RO3003 series) rather then the current glass-reinforced epoxy material (FR-4), with the former having been demonstrated to reach vacuum levels as low as $\sim 10^{-12}$~mbar \cite{sigg2006pcb} and shown to be well suitable for potential future cryogenic operation \cite{RANJAN201587}.  

The field electrodes, which face toward the beam axis, were implemented using four dedicated PCB boards for mimicking quadrupole fields, hereafter referred to as \emph{electrode boards}. These boards carry gold-plated copper segmented pads covering the entire board perimeter using the so-called `round-edge plating` technique, see Fig.~\ref{fig:buncher}a. Electrical connections and mechanical referencing between the \emph{component} and \emph{electrode boards} were achieved using $1$-mm-thick press-fit through-hole pins inserted into a matching via footprint. These define the precise placement of field-electrode segments along the axis and provide low-inductance connections between the passive RF-DC network and the field electrodes. 

For the ``stacked'' PCB-RFQ geometry tested experimentally in this work, shown in Fig.~\ref{fig:buncher}c and Fig.~\ref{fig:buncher}d, 1.6-mm-thick field electrodes were selected for simplicity. The two opposing PCB assemblies were mounted using insulating PEEK spacers, the physical dimensions of which were chosen to achieve an effective field radius of r$_0=3.13$~mm with an aspect ratio of $1:1$. When needed, the entire length between the two \emph{component boards} was covered by a full spacer block to allow for local encapsulation of the PCB-RFQ and, thus, differential pumping, see Fig.~\ref{fig:buncher}d.  

The PCB-RFQ was driven by a custom-made resonant LC circuit producing a matched and opposite-phase pair of sine-wave RF signals \cite{brown2024critical}. To eliminate on-axis RF oscillation and to prevent artificially inflating the ions' energy spread, the amplitudes of the two RF phases were matched to within approximately $ 1\% $. All DC potentials were supplied from high-voltage power supplies (ISEG High Voltage, NHS series).
For the encapsulated PCB-RFQ version, helium buffer gas (Helium 5.0 CP) was injected via a motorized needle valve (Pfeiffer EVR 116) directly into the PCB-RFQ volume.

It should be noted that following the first successful experimental tests of our ``stacked'' PCB-RFQs, multiple design iteration have been produced to provide a larger beam acceptance, which further promotes the application of the PCB-RFQs as a broad quadrupole ion guide and mass filter. This subsequent ``stacked'' PCB-RFQ design iterations have been built with an increased effective field radius of r$_0=5$~mm using 3.2-mm-thick field electrodes.

\section{Performance Characterization of the PCB-RFQ}

\subsection{Ion transport confirmation}

\begin{figure}[tb!]
\centering
\includegraphics[width=1\columnwidth]{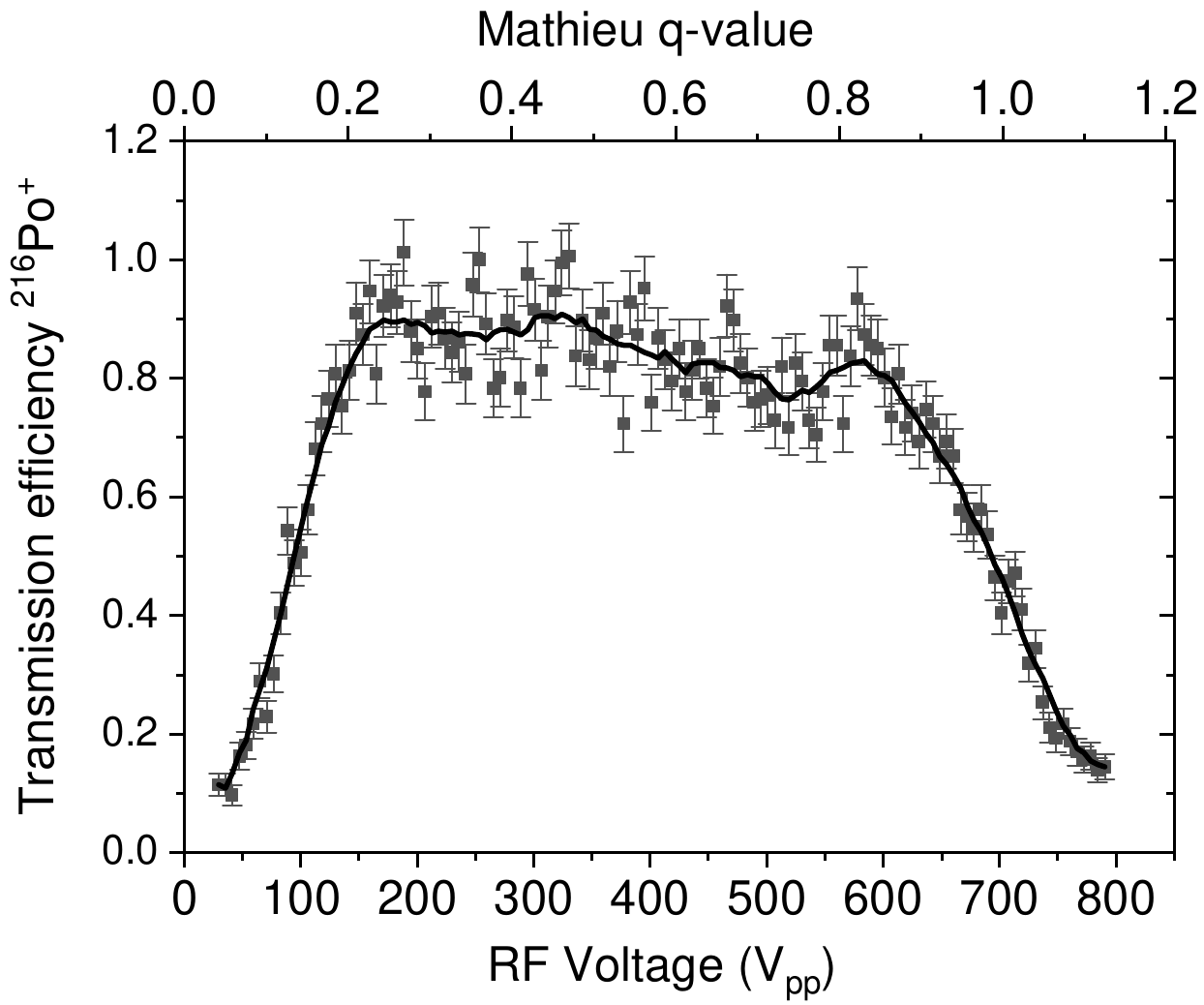}
\caption{Experimental DC-transmission test of our ``stacked'' PCB-RFQs using $^{216}$Po$^+$ ions from a stopping cell identified and counted via $\alpha$ spectroscopy. The line indicates a 10-point running average to guide the eye. Across the transmission plateau an average transport efficiency of $80-90\%$ is reached.}\label{fig:Po_transmission}
\end{figure}

To characterize the performance of our ``stacked'' PCB-RFQ design, first we assessed its capability to transport ions efficiently. A non-encapsulated PCB-RFQ, as shown in Fig.~\ref{fig:buncher}c, was mounted behind a small room-temperature stopping cell, which provided $^{216}$Po$^+$ ions harvested via the $^{228}$Th decay chain. Being an $\alpha$-decaying isotope, $^{216}$Po can be well identified and counted using a Si detector (Hamamatsu S5390) with a quantum efficiency of nearly unity (solid angle $\approx 50\%$ in our case). The stopping cell was filled with helium buffer gas at a pressure of $50$~mbar, which resulted in a helium base pressure of $1.5 \cdot 10^{-2}$~mbar in the first vacuum chamber immediately downstream of the stopping cell. The beam intensity of $^{216}$Po$^+$ ions extracted from the stopping cell was measured by mounting the Si detector directly after the stopping cell. Then the non-encapsulated PCB-RFQ was installed behind the stopping cell and the Si detector was placed after the PCB-RFQ. Fig.~\ref{fig:Po_transmission} shows the $^{216}$Po $\alpha$-decay rate recorded by the Si detector applying different RF amplitudes to the PCB-RFQ (RF amplitudes labelled as V$_\textrm{pp}$ referred to peak-to-peak RF amplitudes). At too low RF amplitudes, the radial confinement is not sufficient to guide the ions all the way to the detector, whereas at too high RF amplitudes the $^{216}$Po$^+$ ions are pushed beyond the low-mass cut-off of the PCB-RFQ. Across a transmission plateau ranging from about $200$~V$_{pp}$ to $600$~V$_{pp}$ (Mathieu q-value $0.2- 0.9$), an average transport efficiency of $80 - 90\%$ is achieved with our ``stacked'' PCB-RFQs. 

\subsection{Quadrupole field quality}
While the transmission plateau behaviour confirms that our PCB-RFQs can be applied as practical ion guides, on its own, it does not provide a measure of their quadrupole field quality. Therefore, mapping the first Mathieu stability region is needed to allow for judging how well our PCB-RFQ acts as an ideal linear quadrupole. 

\begin{figure}[tb!]
\centering
\includegraphics[width=\columnwidth]{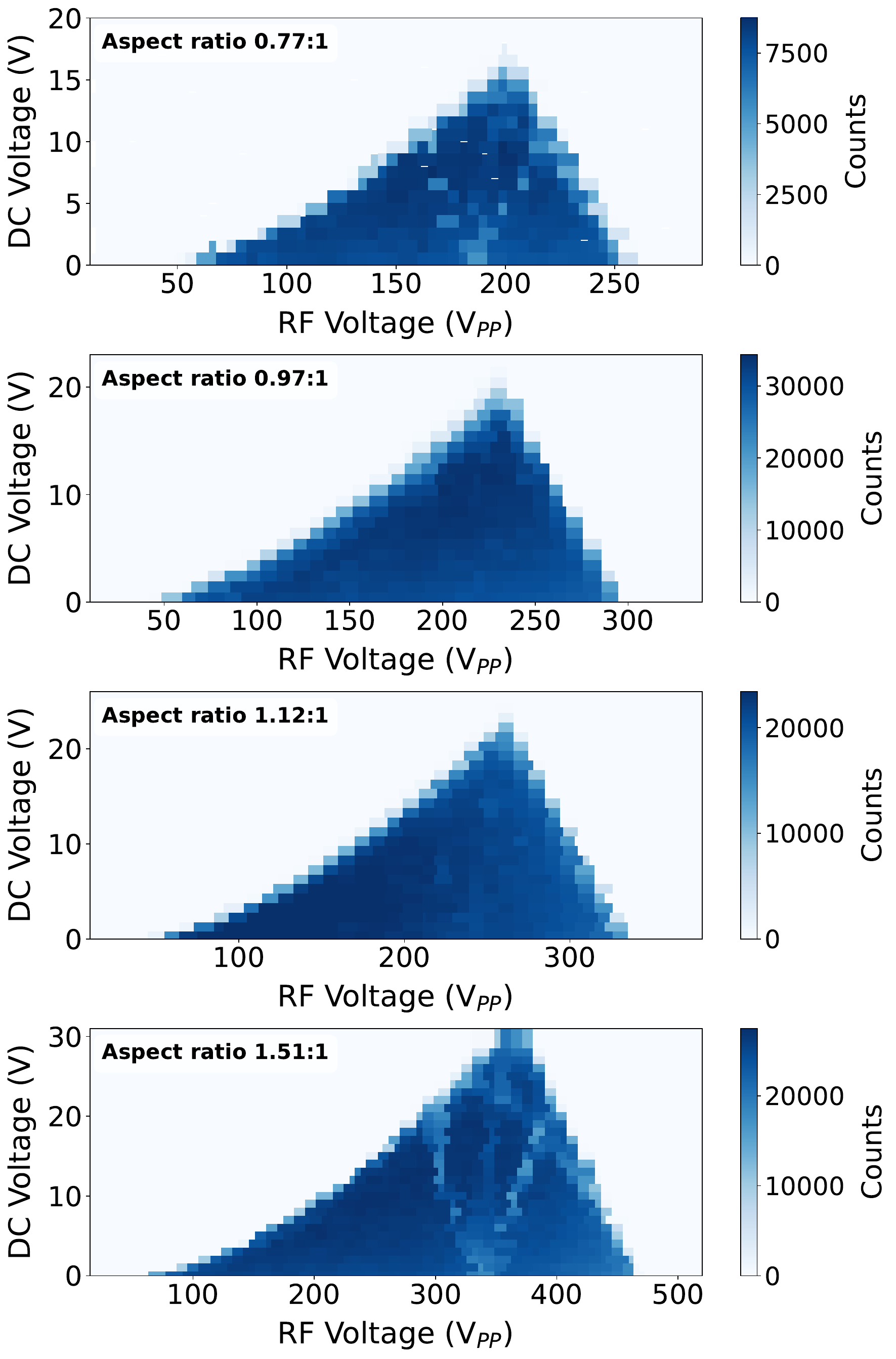}
\caption{Experimental validation of the first Mathieu stability region of different ``stacked'' PCB-RFQ geometries with varying aspect ratios measured with $^{133}$Cs$^+$ ions. The white bands of transmission loss confirm the presence of non-linear resonances caused by the emergence of higher-order multipole components for non-ideal aspect ratios.}
\label{fig:stability}
\end{figure}

To achieve this, the PCB-RFQ setup was modified, enabling the application of a separate (mass filter) DC offset between the RF+ and RF- rails, which enables operating the PCB-RFQ at non-zero Mathieu a-values. The modified PCB-RFQ was then installed in a vacuum chamber, positioned between a thermal ion source delivering $^{133}$Cs$^+$ ions (Heatwaves) and a single-ion-counting detector (ETP/Adaptas, MagenX Micro). The residual gas pressure in the vacuum chamber was below $5 \cdot 10^{-7}$~mbar to rule out effects caused by ion-gas collisions. The first Mathieu stability region was mapped by varying the RF and DC voltages, as shown in Fig.~\ref{fig:stability}. 

In total five different ``stacked'' PCB-RFQ geometries were built and investigated by adjusting the horizontal spacers between the PCB boards, yielding aspect ratios of $ 0.77 $, $ 0.97 $, $ 1.12 $, $ 1.31 $, $ 1.51 : 1 $ for the PCB-RFQ assembly. For the tall and wide PCB-RFQ geometries with aspect ratios of $0.77:1$, $1.31:1$ and $1.51:1$, the emergence of non-linear resonant excitations and corresponding transmission losses was clearly observed in our experimental tests. (Note: the stability region measured with the $1.31:1$ geometry is not shown here [Fig.~\ref{fig:stability}] for brevity, as its measured resonant structure followed the same pattern as the $ 1.51:1 $ geometry.) 

On the other hand, our measurements performed using the PCB-RFQ assemblies with aspect ratios of $ 0.97:1 $ and $ 1.12:1 $ did not exhibit signatures of non-linear resonant extraction, showing that these assemblies mimic the operation of a traditional RFQ cooler-buncher well. 
The experimental behaviour is further consistent with our electrostatic modelling of the higher-order spatial harmonics, which had indicated some robustness of the $ C_{6} $, $ C_{10} $, $ C_{14} $ and $ C_{18} $ coefficients for a range of aspect ratios roughly between $ 0.88$ and $ 1.15$ (\cref{sec:simulations}). Throughout this range of aspect ratios no resonant transmission losses were observed experimentally, which indicates that, overall, the ``stacked'' PCB-RFQ geometry developed herein is somewhat forgiving to dimensional variations, establishing its mechanical robustness for general ion-bunching applications.

\subsection{Cooling and bunching dynamics}
The PCB-RFQ boards were then encapsulated, as shown in Fig.~\ref{fig:buncher}d, using solid PEEK spacer blocks along the board edge to enable differential pumping. This allowed for injecting helium buffer gas into the localized volume inside the PCB-RFQ with typical pressures of $10^{-2}$ to $10^{-3}$~mbar, while maintaining the surrounding vacuum chamber below a pressure of $10^{-5}$~mbar. The local encapsulation allowed us to operate the PCB-RFQ as a dedicated cooler-buncher without the need to use separate vacuum chambers for the cooling-bunching and field-free drift regions. 

For our PCB-RFQ cooler-buncher, a two-stage trapping operation mode \cite{Dickel:200291,SCHWARZ2016131,BARQUEST2016207,REITER2021165823,LECHNER2024169471} was selected to achieve performance characteristics similar to commonly used MR-TOF-MS injection trap systems \cite{DICKEL2015172,Dickel:200291,REITER2021165823}. The first segment of the PCB-RFQ was biased to $1$~kV and a DC gradient of around $0.7-1$~V/cm was used to guide the ions to a local potential minimum at the 40th (final) segment of the PCB-RFQ, which acted as a pre-trap. There, the continuous beam of $^{133}$Cs$^+$ ions was accumulated for $10$~ms and, then, the ions were transferred to an extra trap segment, which acted as the injection trap. The one-segment-long PCB-based injection trap was installed after the main PCB-RFQ board and terminated on both sides by $1$-mm-thick apertures with a hole diameter of $2$~mm. These apertures allowed for precise control of the axial trapping potentials and ensured that the extraction field gradient is as homogeneous as possible when ejecting the ions from the final trap.
Based on simulations, we estimate the maximal non-linearity of this layout to be $ < 2.5 \% $ across the central $\pm 1$~mm of the final trapping segment.

Pre-cooled ions were transferred from the pre-trap to the injection trap, by switching the aperture between the two traps from its confining state (held $30$~V above the voltage applied to the pre‑trap segment) to a low potential ($\pm 1$~V around the central trap potential) for about $5$~$\mu$s. Ions were then cooled again for a variable (re-)cooling time. 
The ion bunches were ejected from the injection trap using a push-pull configuration, accelerated into a drift tube (biased to a negative potential in the range of about $-1$ to $-2$~kV) and guided onto the single-ion-counting detector. Residual RF‑induced perturbations during ion ejection were suppressed by phase‑locking the RF drive to the ejection pulse. 
Detector signals were recorded using either an oscilloscope (Rigol MSO5354) or a time-to-digital converter (FastComtec MCS8) and the ion detection rate was kept below one ion per cycle.  

To evaluate the cooling performance of our PCB-RFQ cooler-buncher, the temporal width of the extracted ion bunch was measured as a function of the cooling time inside the injection trap segment. Measurements were performed over a range of gas flow settings.  
The width of the time-of-flight peaks
followed an exponential decay directly mapping the thermalization of the $^{133}$Cs$^+$ ions with the room-temperature helium buffer gas. 
At the highest flow setting (EVR116 control voltage $4$~V) a cooling time of $ t_{\textrm{cool},1/2} = 34 \pm 3\ \mu\text{s} $ could be achieved, 
which, based on the known cooling cross section of $^{133}$Cs$^+$ ions with helium \cite{DEHMELT196853,Major:1968zz,Moriwaki_1992}, yields an effective He gas pressure of around $\sim 3 \cdot 10^{-2}$~mbar. 
At the lowest flow setting (EVR116 control voltage $3.1$~V) a cooling time of $ t_{\textrm{cool},1/2} = 412 \pm 23\ \mu\text{s} $, corresponding to a local pressure of $\sim 2 \cdot 10^{-3}$~mbar, was determined. In this setting, the ion cloud reached an effective thermal equilibrium after around `$ 4\text{ ms} $` of cooling time. 

\subsection{Ion bunch characterisation}

\begin{figure}[tb]
\centering
\includegraphics[width=1\columnwidth]{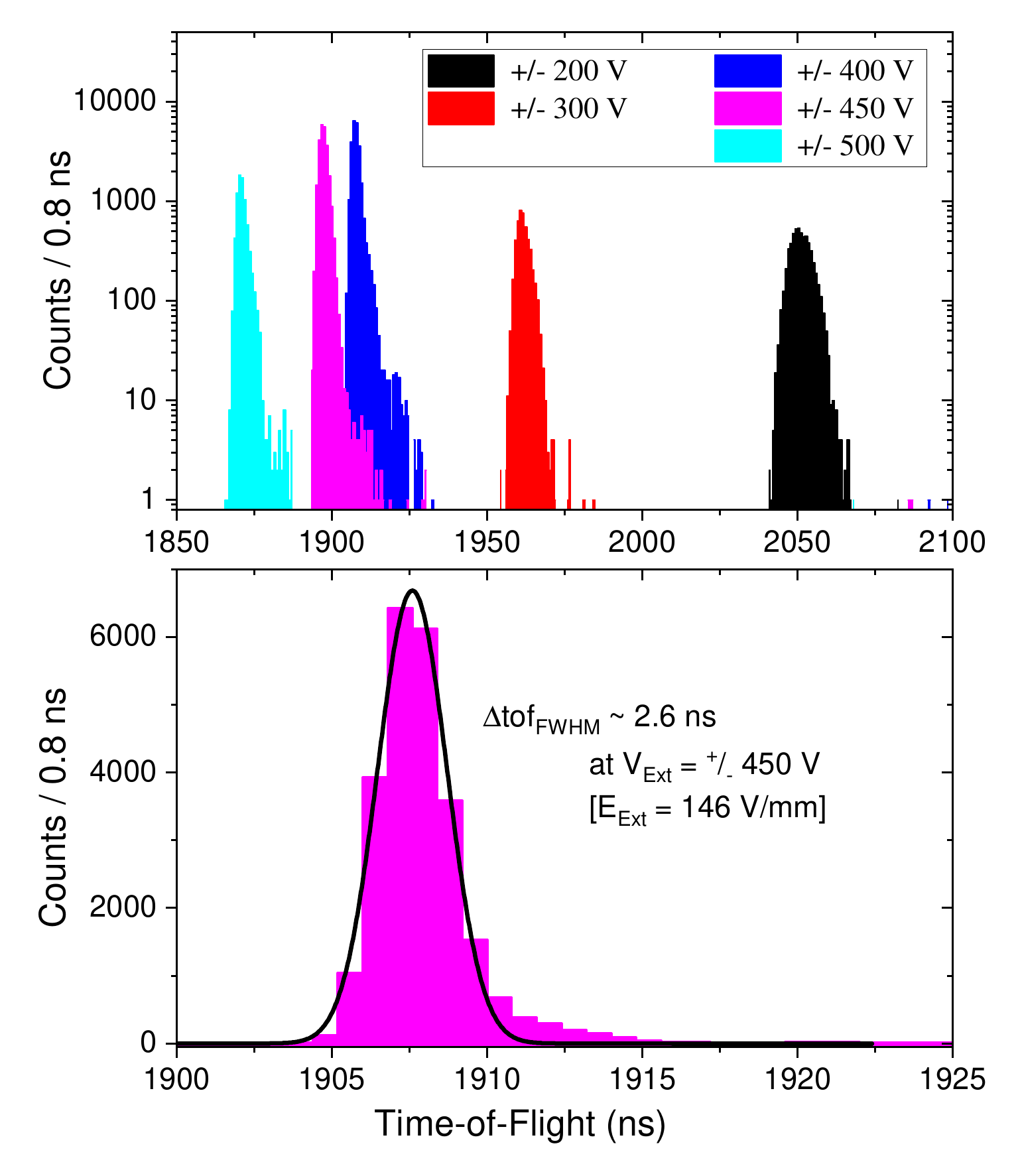}
\caption{
Example time-of-flight spectra of $^{133}$Cs$^+$ ions obtained from the PCB-RFQ cooler-buncher. The top panel shows the ``raw'' time-of-flight spectra, plotted on a logarithmic scale, recorded for different extraction field strengths ($E_\mathrm{ext}$) with the drift-tube voltage held at $-2200$~V. The bottom panel shows same data on a linear scale, featuring the time-of-flight peak achieved with the smallest temporal width. A Gaussian fit to the time‑of‑flight profile (solid black line) yields a minimal width of $\Delta t_\mathrm{FWHM}=2.6\pm0.3$~ns.
}\label{fig:tof_peak}
\end{figure}

Once ions were confirmed to have reached their thermal equilibrium, the ion bunching capabilities of the PCB-RFQ were optimized. 
Fig.~\ref{fig:tof_peak}(top) shows several example time-of-flight spectra of $^{133}$Cs$^+$ ions measured applying different extraction fields. In general, the time-of-flight distributions produced by our PCB-RFQ cooler-buncher can be adequately described by a Gaussian line shape down to a few percent level, at which point peak tails towards longer flight times become visible, particularly at higher extraction field strengths.

\begin{figure}[tb]
\centering
\includegraphics[width=1\columnwidth]{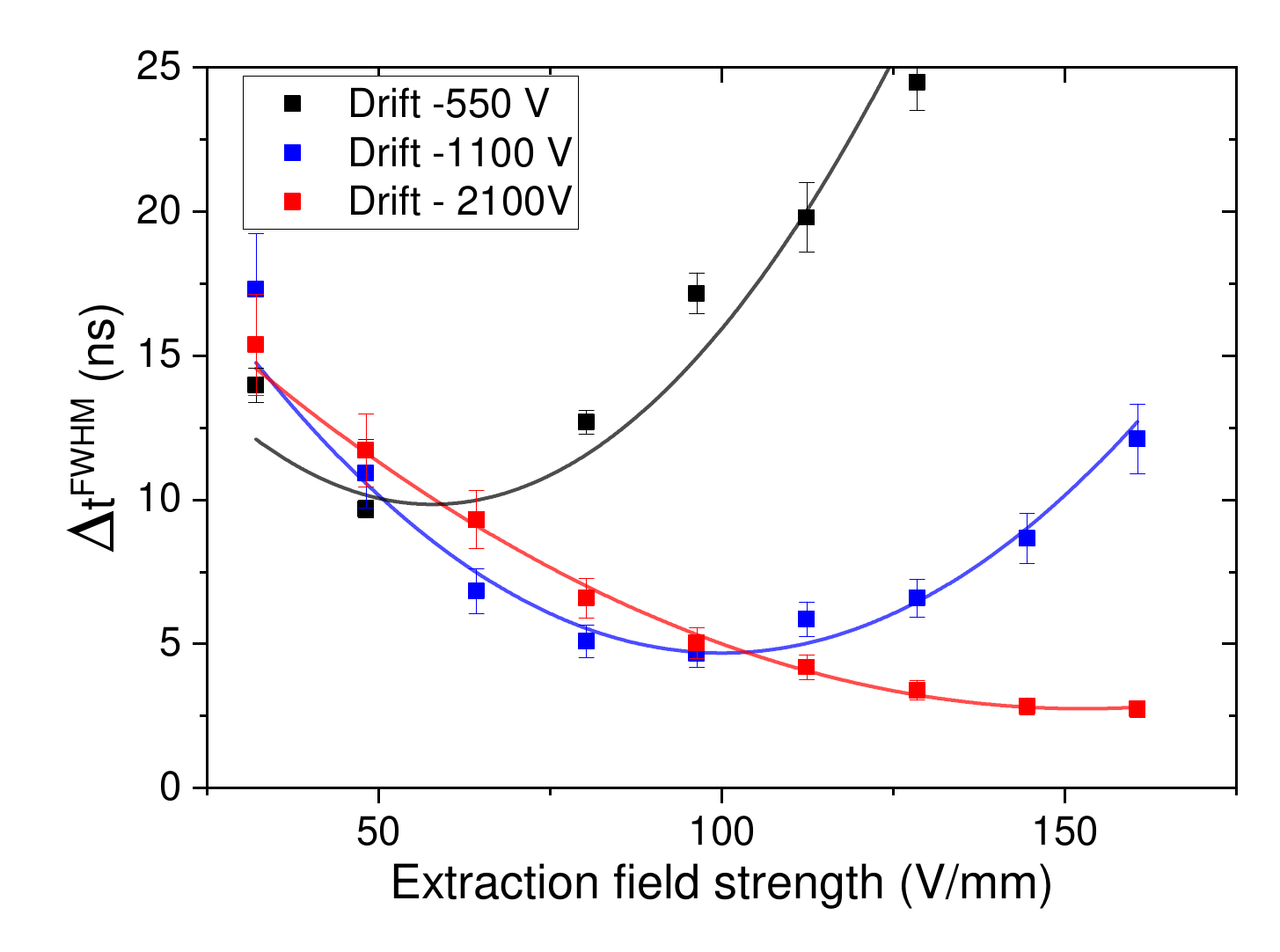}
\caption{Example variation of the time-of-flight peak width measured for $^{133}$Cs$^+$ ions applying different drift-tube voltages as a function of the extraction field strength. The solid lines represent empirical second-order polynomial fits to the data, provided for visual guidance.}
\label{fig:field_scan}
\end{figure}

As only the correct combination of drift-tube voltage and extraction field gradient aligns the first-order time-of-flight focus with the detector \cite{10.1063/1.1715212,yavor2009optics},  Fig.~\ref{fig:field_scan} shows the measured time-of-flight peak widths for different drift-tube voltages of $-550$~V, $-1100$~V and $-2100$~V as a function of the extraction field strength. For these drift voltages, the smallest peak widths were obtained at extraction field strengths of about $50$~V/mm, $100$~V/mm and $150$~V/mm, respectively. The achieved minimum peak width was $\Delta t_\mathrm{FWHM}\approx 2.8$~ns at the drift-tube voltage of $-2100$~V and extraction field strength of $\sim 150$~V/mm. This is about factors of $1.8$ and $3.7$ more narrow compared to the minima obtained at weaker extraction fields of $\sim 100$~V/mm ($\Delta t_\mathrm{FWHM}\approx 5$~ns) and $\sim 50$~V/mm ($\Delta t_\mathrm{FWHM}\approx 10$~ns), respectively, due to the reduced turnaround time in the injection trap.  

Time-of-flight spectra of $^{133}$Cs$^+$ ions were recorded for a range of extraction field gradients using different trapping configurations.
The resulting peak widths at the time focus are shown in \cref{fig:emittance} (labelled as Exp.~1 and Exp.~2). 
At high extraction field strengths sub-$ 3\text{-ns} $ time-of-flight peaks were routinely observed. 
The smallest time-of-flight peak width was found at a push-pull extraction voltage of $\pm 450$~V, corresponding to an extraction field strength of $146$~V/mm, and drift tube voltage of $-2200$~V. A Gaussian fit to the time-of-flight profile confirmed the narrow temporal width as $ \Delta t_{\text{FWHM}} = 2.6 \pm 0.3\ \text{ns} $, as shown in Fig.~\ref{fig:tof_peak}(bottom). 
At this point, the time-of-flight peak width approached the width of the single-ion-response signal of the MagneX micro detector, which, in our case, corresponded to $\sim 2.5$~ns at a trigger level of $5$~mV. In the present study, the time-of-flight peak width could not be further reduced.  

In order to judge if such sub-$ 3\text{-ns} $ ion distributions would fall within the acceptance of common MR-TOF-MS devices of $\Delta K / K \sim \text{few}\ \%$, one has to also quantify the ions' energy spread.  

\begin{figure}[tb]
\centering
\includegraphics[width=\columnwidth]{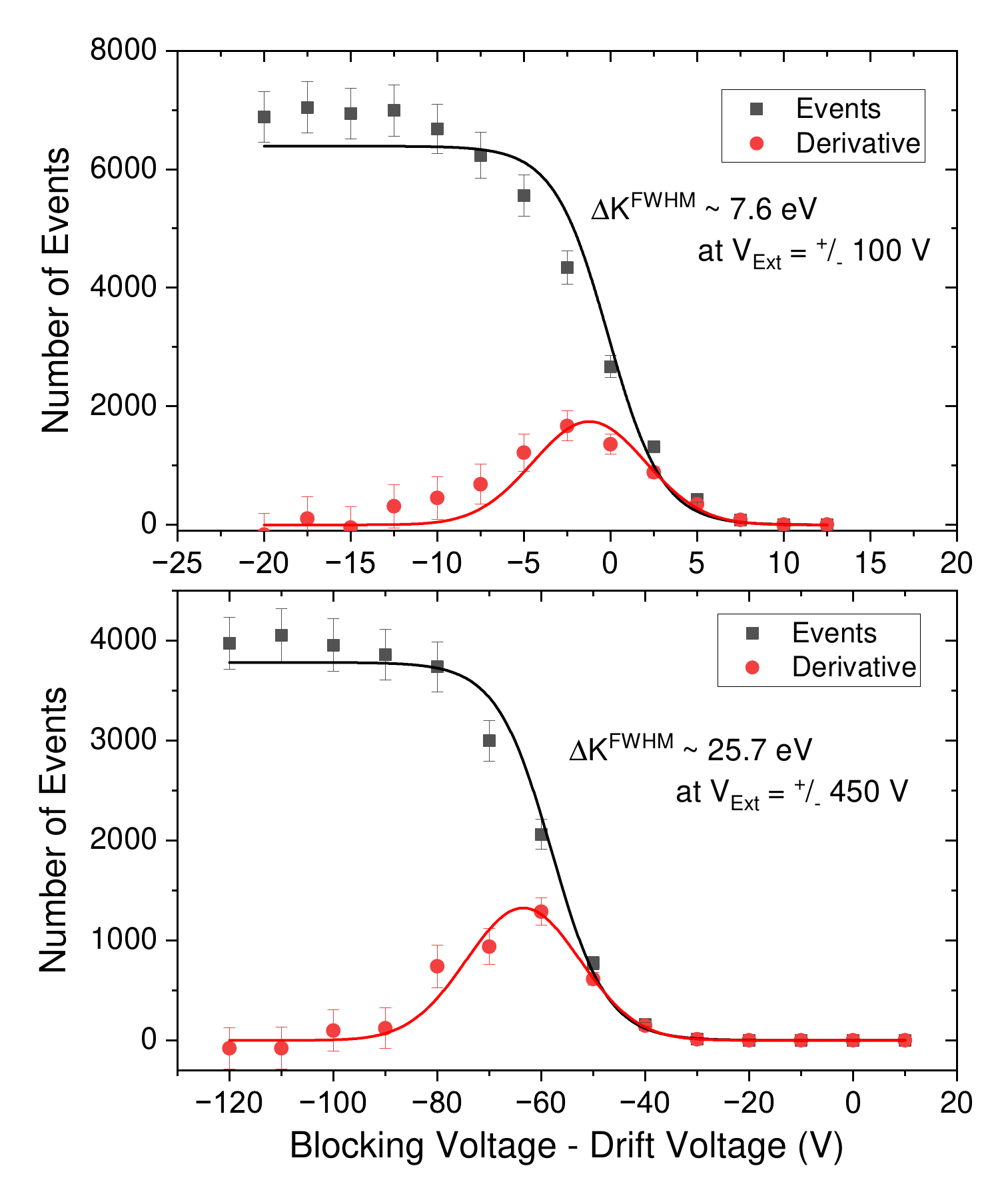}
\caption{Example energy-spread measurements using a self-made Retarding Field Analyzer (RFA) for extraction voltages of $V_\mathrm{ext}= \pm 100$~V (top) and $\pm 450$~V (bottom) resulting in energy spreads (FWHM) of $7.6$ and $25.7$~eV, respectively. The solid lines represent fits to the derivative of the intensity response with a Gaussian line shape.}
\label{fig:energy}
\end{figure}

To determine the energy spread, a self-made Retarding Field Analyzer (RFA) consisting of five consecutive $79\%$-transmission stainless-steel meshes (TWP Inc. 100 Mesh T316), separated by alternating $1$-mm-thick PTFE and stainless-steel spacers, was installed after the drift tube (labelled as Exp.~3 in \cref{fig:emittance}). In our RFA unit, the two outermost meshes were biased to the same negative potential as the drift tube, whereas the central mesh was used at a variable blocking potential. The two remaining neighboring meshes were used to reduce field penetration effects and were held at a voltage of about $80$~V lower than the required blocking voltage. The intensity of the $^{133}$Cs$^+$ ion beam was recorded while incrementally sweeping the blocking potentials of the RFA. The energy spread of the ion bunch could then be obtained by fitting the derivative of the intensity response with a Gaussian line shape. Fig.~\ref{fig:energy} shows two example blocking-potential scans for two different push-pull extraction voltage settings ($\pm 100$~V and $\pm 450$~V) and corresponding extraction field strengths ($32$~V/mm and $146$~V/mm). For the moderate extraction field strength of $32$~V/mm, an energy spread of $\Delta K^{\textrm{FWHM}} = 7.6 \pm 1$~eV was measured, whereas for the strong $146$~V/mm extraction field settings, the RFA scan revealed an energy spread of $25.7 \pm 1$~eV. 

A range of RFA scans and corresponding time-of-flight peak width measurements were performed for different extraction field gradients (Exp.~3), with the results shown in Fig.~\ref{fig:emittance}. Particularly at weak extraction fields, the energy-spread measurements were limited by effects due to field penetration, which were estimated to contribute $ 1.3 \pm 0.3\ \text{eV} $ to the measurement floor. The lowest energy spread of an ion bunch recorded in this work, was $ \Delta K^{\textrm{FWHM}} = 2.5 \pm 1\ \text{eV} $.

\subsection{Longitudinal emittance determination}
\begin{figure}[tb!]
\centering
\includegraphics[width=1\columnwidth]{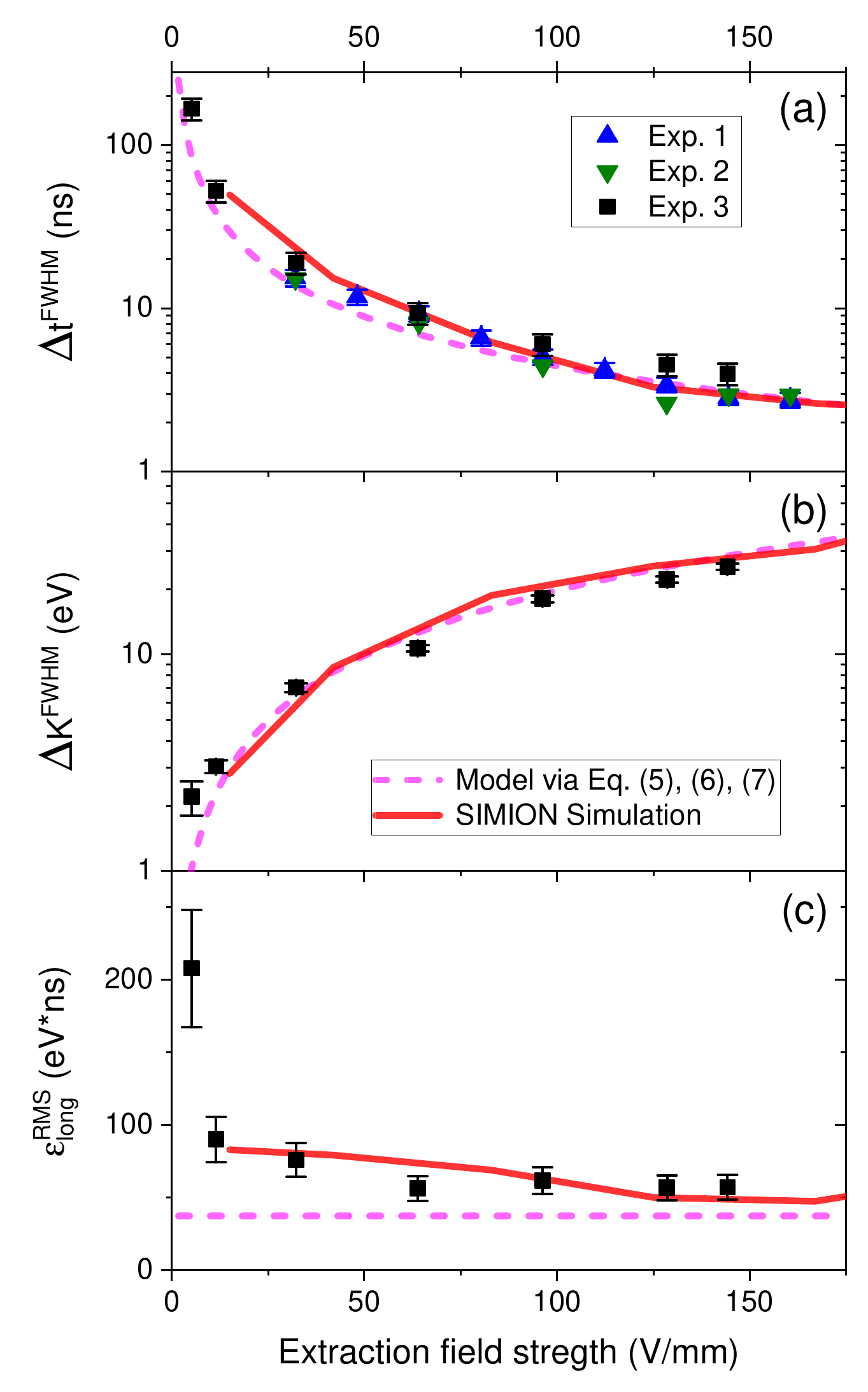}
\caption{Systematic study comparing (a) the time-of-flight peak widths, (b) the ions' energy spread, (c) and the resulting longitudinal emittance of the produced ion bunches for different extraction fields. The experimental data points are compared with results from SIMION simulations (solid red lines) and analytical predictions based on \cref{eq:DK,eq:DT,eq:emmittance} (dashed magenta lines). The data were measured using different trapping configurations; while Exp.~1 and Exp.~2 was performed without the RFA, Exp.~3 was performed with the RFA mounted in the vacuum chamber allowing for a simultaneous measurement of the energy and temporal spreads under identical conditions, as shown in all three panels; (a), (b) and (c). The resulting longitudinal emittance is almost independent of the extraction field strength.}
\label{fig:emittance}
\end{figure}

Having experimentally measured both the temporal and energy spreads of the ion bunches extracted from our PCB-RFQ cooler-buncher allows for the calculation of its longitudinal emittance using \cref{eq:emmittance} with the results shown in Fig.~\ref{fig:emittance}c. As expected, over a wide range of extraction field settings, the longitudinal emittance of the system is well conserved. Considering only the settings with $E_\mathrm{ext}>20$~V/mm yields a longitudinal emittance of $ 58 \pm 4\ \text{eV}\cdot\text{ns} $. 

To better understand the limitations and experimental performance of the new PCB-RFQ cooler-buncher, we compared its capabilities to predictions based on the analytical models discussed in \cref{RFQs} and to a dedicated SIMION bunching simulation. We used \cref{eq:DK,eq:DT,eq:emmittance} to calculate the expected minimum temporal and energy spreads and the corresponding longitudinal emittance of the ion bunches, as shown by the dashed magenta lines in Fig.~\ref{fig:emittance}. Assuming a thermal ion motion at an equilibrium temperature of $300$~K and a spatial ion cloud size (FWHM) along the beam axis prior to ejection of $0.2$~mm (obtained from SIMION simulations), the analytical model describes the bunching performance of our PCB RFQ well. 
Our SIMION bunching simulations, as shown by the solid red lines in Fig.~\ref{fig:emittance}, are also in good agreement with the observed experimental performance. 
Overall, the calculated longitudinal emittance based on both the analytical model and the simulations agrees well with the experimentally obtained emittance.   

The excellent phase-space compression confirms the strength of our PCB-RFQ design and is a direct consequence of its deep axial confinement, matched by the radial pseudopotential. A longitudinal emittance of $ 58 \pm 4\ \text{eV}\cdot\text{ns} $ compares favourably (improvement of about a factor of $2-4$) with other state-of-the-art specialized ion-beam-preparation devices, including legacy cooler-bunchers and even proposed next-generation cooler-bunchers \cite{VIRTANEN2025170186,BARQUEST2016207,PODADERAALISEDA2004647,LECHNER2024169471,Varentsov_2025}, and ion traps operated as dedicated injectors for contemporary MR-TOF-MS systems \cite{Dickel:200291,ITO2013544,atoms11110139}. Because the emittance is inherently small, our PCB-RFQ setup can apply weak extraction gradients to achieve either sub-`$ 3\text{-eV} $` energy spreads as required, e.g., for collinear laser spectroscopy, or apply high extraction gradients to reach sub-`$ 3\text{-ns} $` temporal spreads well suited for high-resolution time-of-flight spectrometers.


\section{Conclusions and Outlook}
We have successfully simulated, developed, and experimentally validated a novel linear PCB-RFQ cooler-buncher utilizing purely printed circuit board technology. Overcoming the characteristic non-linear field phenomena associated with basic planar (flat) electrodes, the implementation of a stacked board geometry successfully mimics the electric fields and pseudopotential characteristics of conventionally machined RFQ traps. 

The integration of passive electronics, such as the DC gradient resistor chain, RF blocking networks, and an active internal heating layer, directly into the PCB substrate and the usage of press-fit pins drastically reduce any alignment tolerance issues, manufacturing lead times, and the physical footprint generally associated with standard rod-based RFQ cooler-bunchers. 

Experimentally, the system delivers highly competitive beam properties. It delivers a compressed longitudinal emittance of just $ 58 \pm 4\ \text{eV}\cdot\text{ns} $. Thus, the PCB-RFQ cooler-buncher may act as a highly flexible interface for RIB facilities and general ion bunching applications. We demonstrate the capacity of our PCB-RFQ cooler-buncher to generate either below $ 3\ \text{eV} $ energy-focused ion bunches appropriate for energy-sensitivity spectroscopy, or $ 2.6\ \text{ns} $ time-focused ion bunches ideal for high-resolution MR-TOF-MS measurements, or almost any combination between those two limits. 

\begin{figure}[tb!]
\centering
\includegraphics[width=0.9\columnwidth]{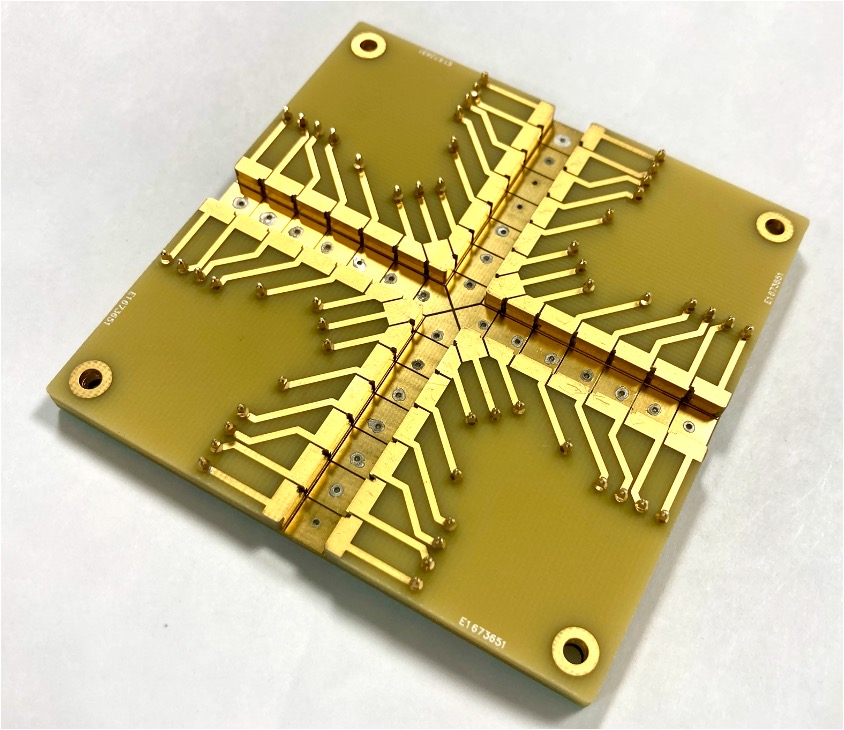}
\caption{Photograph of the novel PCB-RFQ switchyard prototype, which supports the seamless merging and splitting of independent ion beams.}
\label{fig:PCB_switchyard}
\end{figure}

Our rapidly scalable, cost-effective stacked PCB architecture establishes a robust blueprint for deploying high-performance trapping, cooling, and bunching hardware across an expanding field of precision physics and spectrometry experiments. Beyond the linear geometries characterized in this work, the surface topology and on-board photolithographic alignment allow for building a wider range of non-traditional, complex ion-optical configurations.
Specifically, this platform provides an elegant route to construct S-shaped or 90-degree-bend RFQs, such as the ones from Refs.~\cite{PhysRevResearch.3.043041,MORIN2026166027}, which, due to their curved geometries, allow for line-of-sight access to the ion beams for beam purification and spectroscopy, while transmitting and cooling the ions of interest. 
Furthermore, our modular architecture also enables the construction of a 4-way PCB-RFQ switchyard, inspired by macroscopic RFQ switchyards such as the ones from Refs.~\cite{Plass_2015,REITER2021165823,YU2024169371}. A prototype of which has already been built, as shown in Fig.~\ref{fig:PCB_switchyard}. Our PCB-RFQ switchyard will act as a highly symmetric, buffer-gas-filled intersection connecting four orthogonal linear RFQ channels at a single point. By leveraging the fine segmentation of the PCB traces to apply localized DC steering potentials, the switchyard will be able to steer an incoming ion package into any of the three other directions and support merging and splitting of independent ion beams \cite{Plass_2015,REITER2021165823,YU2024169371}. At contemporary RIB facilities, this will allow for, e.g., the continuous co-injection of stable, offline calibration ions alongside the precious radioactive ions. 

Ultimately, the stacked PCB-RFQ platform bridges the gap between complex classical ion optics and standard, commercial microfabrication, offering a versatile toolkit for next-generation precision beam preparation and steering.

\section*{Acknowledgements}
The authors thank the technical staff at the University of Edinburgh for their support in designing and populating the PCB boards. This work was supported by the Science and Technology Facilities Council (STFC, UK) under grant agreements no. ST/X004953/1, ST/V001051/1 and ST/Y000293/1, by the German Federal Ministry of Research, Technology and Space (BMFTR) under grant agreements no.\ 05P21RGFN1 and 05P24RG4, by the German Research Foundation (DFG) under grant agreements no.\ 422761894 and AY 155/2-1, by HGS-HIRe and by Justus-Liebig-Universit{\"a}t Gie{\ss}en and GSI under the JLU-GSI strategic Helmholtz partnership agreement.

\bibliography{bibli}

@article{PhysRevLett.88.094801,
  title = {{On-Line Ion Cooling and Bunching for Collinear Laser Spectroscopy}},
  author = {Nieminen, A. and Campbell, P. and Billowes, J. and Forest, D. H. and Griffith, J. A. R. and Huikari, J. and Jokinen, A. and Moore, I. D. and Moore, R. and Tungate, G. and \"Ayst\"o, J.},
  journal = {Phys. Rev. Lett.},
  volume = {88},
  issue = {9},
  pages = {094801},
  numpages = {4},
  year = {2002},
  month = {Feb},
  publisher = {American Physical Society},
  doi = {10.1103/PhysRevLett.88.094801},
  url = {https://link.aps.org/doi/10.1103/PhysRevLett.88.094801}
}

@article{YANG2023104005,
	author = {X.F. Yang and S.J. Wang and S.G. Wilkins and R.F. Garcia Ruiz},
	doi = {https://doi.org/10.1016/j.ppnp.2022.104005},
	issn = {0146-6410},
	journal = {Progress in Particle and Nuclear Physics},
	pages = {104005},
	title = {Laser spectroscopy for the study of exotic nuclei},
	url = {https://www.sciencedirect.com/science/article/pii/S0146641022000631},
	volume = {129},
	year = {2023}}

@article{WOLF201282,
	author = {R.N. Wolf and D. Beck and K. Blaum and Ch. B{\"o}hm and Ch. Borgmann and M. Breitenfeldt and F. Herfurth and A. Herlert and M. Kowalska and S. Kreim and D. Lunney and S. Naimi and D. Neidherr and M. Rosenbusch and L. Schweikhard and J. Stanja and F. Wienholtz and K. Zuber},
	doi = {https://doi.org/10.1016/j.nima.2012.05.067},
	issn = {0168-9002},
	journal = {Nuclear Instruments and Methods in Physics Research Section A: Accelerators, Spectrometers, Detectors and Associated Equipment},
	pages = {82-90},
	title = {On-line separation of short-lived nuclei by a multi-reflection time-of-flight device},
	url = {https://www.sciencedirect.com/science/article/pii/S016890021200575X},
	volume = {686},
	year = {2012}}

@article{ROSENBUSCH2023167824,
	author = {M. Rosenbusch and M. Wada and S. Chen and A. Takamine and S. Iimura and D. Hou and W. Xian and S. Yan and P. Schury and Y. Hirayama and Y. Ito and H. Ishiyama and S. Kimura and T. Kojima and J. Lee and J. Liu and S. Michimasa and H. Miyatake and J.Y. Moon and M. Mukai and S. Naimi and S. Nishimura and T. Niwase and T. Sonoda and Y.X. Watanabe and H. Wollnik},
	doi = {https://doi.org/10.1016/j.nima.2022.167824},
	issn = {0168-9002},
	journal = {Nuclear Instruments and Methods in Physics Research Section A: Accelerators, Spectrometers, Detectors and Associated Equipment},
	pages = {167824},
	title = {The new MRTOF mass spectrograph following the ZeroDegree spectrometer at RIKEN's RIBF facility},
	url = {https://www.sciencedirect.com/science/article/pii/S0168900222011160},
	volume = {1047},
	year = {2023}}

@Article{Cooper2019,
author={Cooper, B. S.
and A.Perrett, H.
and Ricketts, C. M.
and Read, C.
and Edwards, G.
and Flanagan, K. T.
and Billowes, J.
and Binnersley, C. L.
and Bissell, M. L.
and Cocolios, T. E.
and de Groote, R. P.
and Farooq-Smith, G. J.
and Ruiz, R. F. Garcia
and Gins, W.
and Koszorus, A.
and Neyens, G.
and P-Gustafsson, F.
and Stroke, H. H.
and Vernon, A. R.
and Wendt, K. D. A.
and Wilkins, S. G.
and Yang, X. F.},
title={{A compact RFQ cooler buncher for CRIS experiments}},
journal={Hyperfine Interact.},
year={2019},
month={May},
day={13},
volume={240},
number={1},
pages={52},
issn={1572-9540},
doi={10.1007/s10751-019-1586-7},
url={https://doi.org/10.1007/s10751-019-1586-7}
}

@article{Moriwaki_1992,
	author = {Moriwaki, Yoshiki and Tachikawa, Maki and Yoshiharu Maeno, Yoshiharu Maeno and Tadao Shimizu, Tadao Shimizu},
	doi = {10.1143/JJAP.31.L1640},
	journal = {Jpn. J. Appl. Phys.},
	month = {nov},
	number = {11B},
	pages = {L1640},
	title = {{Collision Cooling of Ions Stored in Quadrupole Radio-Frequency Trap}},
	url = {https://doi.org/10.1143/JJAP.31.L1640},
	volume = {31},
	year = {1992}}

@article{Major:1968zz,
    author = "Major, F. G. and Dehmelt, H. G.",
    title = "{Exchange-Collision Technique for the rf Spectroscopy of Stored Ions}",
    doi = "10.1103/PhysRev.170.91",
    journal = "Phys. Rev.",
    volume = "170",
    pages = "91--107",
    year = "1968"
}

@article{Decker:2019aa,
	author = {Decker, Trevor K. and Zheng, Yajun and Ruben, Aaron J. and Wang, Xiao and Lammert, Stephen A. and Austin, Daniel E. and Hawkins, Aaron R.},
	date = {2019/03/01},
	doi = {10.1007/s13361-018-2104-x},
	id = {Decker2019},
	isbn = {1879-1123},
	journal = {J. Am. Soc. Mass Spectrom.},
	number = {3},
	pages = {482--488},
	title = {{A Microscale Planar Linear Ion Trap Mass Spectrometer}},
	url = {https://doi.org/10.1007/s13361-018-2104-x},
	volume = {30},
	year = {2019}}

@article{CHENG2024100364,
	author = {Yu-Peng Cheng and Chen Shen and Zhang-Xu Wu and Shan Li and You-Jiang Liu and Han Wang and Chi-Lai Chen},
	doi = {https://doi.org/10.1016/j.cjac.2024.100364},
	issn = {1872-2040},
	journal = {Chin. J. Anal. Chem.},
	number = {2},
	pages = {100364},
	title = {{Simulation study of a planar quadrupole mass filter for MEMS mass spectrometer}},
	url = {https://www.sciencedirect.com/science/article/pii/S1872204024000094},
	volume = {52},
	year = {2024}}

@article{SONG2006631,
	author = {Yishu Song and Guangxiang Wu and Qingyu Song and R. Graham Cooks and Zheng Ouyang and Wolfgang R. Plass},
	doi = {https://doi.org/10.1016/j.jasms.2005.12.014},
	issn = {1044-0305},
	journal = {J. Am. Soc. Mass Spectrom.},
	number = {4},
	pages = {631-639},
	title = {{Novel Linear Ion Trap Mass Analyzer Composed of Four Planar Electrodes}},
	url = {https://www.sciencedirect.com/science/article/pii/S104403050600002X},
	volume = {17},
	year = {2006}}

@techreport{sigg2006pcb,
  author      = {Sigg, Daniel},
  title       = {{Printed Circuit Boards for Ultra High Vacuum}},
  institution = {LIGO Laboratory},
  year        = {2006},
  number      = {{LIGO-T060280-00}},
  type        = {Technical Note},
  url         = {https://ific.uv.es/elec/files/PCB-UHV.pdf}
}

@article{gp2b-krwb,
  title = {Commissioning and full realization of a collinear resonance ionization spectroscopy setup at Beijing Radioactive Ion-beam Facility},
  author = {Mei, W. C. and Hu, H. R. and Guo, Y. F. and Yan, Z. and Yang, X. F. and Chen, S. J. and Chen, D. Y. and Lin, Y. P. and Liu, Y. S. and Zhang, C. and Jing, Y. P. and Gao, T. X. and Shen, X. and Jia, Y. Y. and Lin, Y. T. and Zhang, H. X. and Bai, S. W. and Tang, B. and Ma, X. and Song, G. F. and Ye, S. and Lu, M. Y. and Dong, J. Y. and Dong, B. K. and Lv, J. H. and Dong, S. Y. and Liu, F. C. and Hu, Z. and Liu, X. and Zhu, S. T. and Yi, Y. L. and He, C. Y. and Takamine, A. and Cui, B. Q. and Yang, J. and Liu, Z. Y. and Su, J. and Liu, H. N. and Ye, Y. L. and Guo, B.},
  journal = {Phys. Rev. Res.},
  volume = {8},
  issue = {2},
  pages = {023286},
  numpages = {10},
  year = {2026},
  month = {Jun},
  publisher = {American Physical Society},
  doi = {10.1103/gp2b-krwb},
  url = {https://link.aps.org/doi/10.1103/gp2b-krwb}
}

@article{GAMAGE2020116344,
	author = {Radhya W. Gamage and Daniel E. Austin},
	doi = {https://doi.org/10.1016/j.ijms.2020.116344},
	issn = {1387-3806},
	journal = {Int. J. Mass Spectrom.},
	pages = {116344},
	title = {{The effects of electrode misalignments on the performance of a miniaturized linear wire ion trap mass spectrometer}},
	url = {https://www.sciencedirect.com/science/article/pii/S1387380620301093},
	volume = {453},
	year = {2020}}

@article{Busch:1961aa,
	author = {v. Busch, F. and Paul, W.},
	date = {1961/10/01},
	doi = {10.1007/BF01378433},
	id = {v. Busch1961},
	isbn = {0044-3328},
	journal = {Zeitschrift f{\"u}r Physik},
	number = {5},
	pages = {588--594},
	title = {{{\"U}ber nichtlineare Resonanzen im elektrischen Massenfilter als Folge von Feldfehlern}},
	url = {https://doi.org/10.1007/BF01378433},
	volume = {164},
	year = {1961}}

@article{FRANZEN199415,
	author = {J. Franzen},
	doi = {https://doi.org/10.1016/0168-1176(93)03907-4},
	issn = {0168-1176},
	journal = {Int. J. Mass Spectrom. Ion Process.},
	number = {1},
	pages = {15-40},
	title = {{The non-linear ion trap. Part 5. Nature of non-linear resonances and resonant ion ejection}},
	url = {https://www.sciencedirect.com/science/article/pii/0168117693039074},
	volume = {130},
	year = {1994}}

@article{HAETTNER2018138,
	author = {Emma Haettner and Wolfgang R. Pla{\ss} and Ulrich Czok and Timo Dickel and Hans Geissel and Wadim Kinsel and Martin Petrick and Thorsten Sch{\"a}fer and Christoph Scheidenberger},
	doi = {https://doi.org/10.1016/j.nima.2017.10.003},
	issn = {0168-9002},
	journal = {Nucl. Instrum. Methods Phys. Res. A},
	pages = {138-151},
	title = {A versatile triple radiofrequency quadrupole system for cooling, mass separation and bunching of exotic nuclei},
	url = {https://www.sciencedirect.com/science/article/pii/S0168900217310409},
	volume = {880},
	year = {2018}}

@article{Tian:2018aa,
	author = {Tian, Yuan and Decker, Trevor K. and McClellan, Joshua S. and Wu, Qinghao and De la Cruz, Abraham and Hawkins, Aaron R. and Austin, Daniel E.},
	date = {2018/07/01},
	doi = {10.1007/s13361-018-1942-x},
	id = {Tian2018},
	isbn = {1879-1123},
	journal = {J. Am. Soc. Mass Spectrom.},
	number = {7},
	pages = {1376--1385},
	title = {{Experimental Observation of the Effects of Translational and Rotational Electrode Misalignment on a Planar Linear Ion Trap Mass Spectrometer}},
	url = {https://doi.org/10.1007/s13361-018-1942-x},
	volume = {29},
	year = {2018}}

@article{Conseil-Gudla:2017aa,
	author = {Conseil-Gudla, H{\'e}l{\`e}ne and Gudla, Visweswara C. and Borgaonkar, Shruti and Jellesen, Morten S. and Ambat, Rajan},
	date = {2017/04/01},
	doi = {10.1007/s10854-016-6292-5},
	id = {Conseil-Gudla2017},
	isbn = {1573-482X},
	journal = {J. Mater. Sci.: Mater. Electron.},
	number = {8},
	pages = {6138--6151},
	title = {{Investigation of moisture uptake into printed circuit board laminate and solder mask materials}},
	url = {https://doi.org/10.1007/s10854-016-6292-5},
	volume = {28},
	year = {2017}}

@article{ITO2013544,
	author = {Y. Ito and P. Schury and M. Wada and S. Naimi and C. Smorra and T. Sonoda and H. Mita and A. Takamine and K. Okada and A. Ozawa and H. Wollnik},
	doi = {https://doi.org/10.1016/j.nimb.2013.07.069},
	issn = {0168-583X},
	journal = {Nucl. Instrum. Methods Phys. Res. B},
	pages = {544-549},
	title = {{A novel ion cooling trap for multi-reflection time-of-flight mass spectrograph}},
	url = {https://www.sciencedirect.com/science/article/pii/S0168583X13008902},
	volume = {317},
	year = {2013}}

@article{Bautista-Salvador_2019,
	author = {Bautista-Salvador, A and Zarantonello, G and Hahn, H and Preciado-Grijalva, A and Morgner, J and Wahnschaffe, M and Ospelkaus, C},
	doi = {10.1088/1367-2630/ab0e46},
	journal = {New Journal of Physics},
	month = {apr},
	number = {4},
	pages = {043011},
	publisher = {IOP Publishing},
	title = {Multilayer ion trap technology for scalable quantum computing and quantum simulation},
	url = {https://doi.org/10.1088/1367-2630/ab0e46},
	volume = {21},
	year = {2019}}

@article{hughes2011microfabricated,
  title={{Microfabricated ion traps}},
  author={Hughes, Marcus D and Lekitsch, Bjoern and Broersma, Jiddu A and Hensinger, Winfried K},
  journal={Contemporary Physics},
  volume={52},
  number={6},
  pages={505--529},
  year={2011},
  publisher={Taylor \& Francis},
  doi = {10.1080/00107514.2011.601918},
  url = {https://doi.org/10.1080/00107514.2011.601918}
}

@article{Schneider2014PRA,
  title = {{Laser-Cooling-Assisted Mass Spectrometry}},
  author = {Schneider, Christian and Schowalter, Steven J. and Chen, Kuang and Sullivan, Scott T. and Hudson, Eric R.},
  journal = {Phys. Rev. Appl.},
  volume = {2},
  issue = {3},
  pages = {034013},
  numpages = {7},
  year = {2014},
  month = {Sep},
  publisher = {American Physical Society},
  doi = {10.1103/PhysRevApplied.2.034013},
  url = {https://link.aps.org/doi/10.1103/PhysRevApplied.2.034013}
}

@article{Zhang:2015aa,
	author = {Zhang, Xinyu and Garimella, Sandilya V. B. and Prost, Spencer A. and Webb, Ian K. and Chen, Tsung-Chi and Tang, Keqi and Tolmachev, Aleksey V. and Norheim, Randolph V. and Baker, Erin S. and Anderson, Gordon A. and Ibrahim, Yehia M. and Smith, Richard D.},
	date = {2015/06/16},
	doi = {10.1021/acs.analchem.5b00214},
	isbn = {0003-2700},
	journal = {Analytical Chemistry},
	journal1 = {Analytical Chemistry},
	journal2 = {Anal. Chem.},
	month = {06},
	number = {12},
	pages = {6010--6016},
	publisher = {American Chemical Society},
	title = {Ion Trapping, Storage, and Ejection in Structures for Lossless Ion Manipulations},
	type = {doi: 10.1021/acs.analchem.5b00214},
	url = {https://doi.org/10.1021/acs.analchem.5b00214},
	volume = {87},
	year = {2015},
	year1 = {2015}}

@article{Li:2009aa,
	author = {Li, Xiaoxu and Jiang, Gongyu and Luo, Chan and Xu, Fuxing and Wang, Yuanyuan and Ding, Li and Ding, Chuan-Fan},
	date = {2009/06/15},
	doi = {10.1021/ac900478e},
	isbn = {0003-2700},
	journal = {Analytical Chemistry},
	journal1 = {Analytical Chemistry},
	journal2 = {Anal. Chem.},
	month = {06},
	number = {12},
	pages = {4840--4846},
	publisher = {American Chemical Society},
	title = {Ion Trap Array Mass Analyzer: Structure and Performance},
	type = {doi: 10.1021/ac900478e},
	url = {https://doi.org/10.1021/ac900478e},
	volume = {81},
	year = {2009},
	year1 = {2009}}

@article{Stick:2006aa,
	author = {Stick, D. and Hensinger, W. K. and Olmschenk, S. and Madsen, M. J. and Schwab, K. and Monroe, C.},
	date = {2006/01/01},
	doi = {10.1038/nphys171},
	id = {Stick2006},
	isbn = {1745-2481},
	journal = {Nature Physics},
	number = {1},
	pages = {36--39},
	title = {Ion trap in a semiconductor chip},
	url = {https://doi.org/10.1038/nphys171},
	volume = {2},
	year = {2006}}

@article{Xu:2025aa,
	author = {Xu, Shuqi and Xia, Xiaoxing and Yu, Qian and Parakh, Abhinav and Khan, Sumanta and Megidish, Eli and You, Bingran and Hemmerling, Boerge and Jayich, Andrew and Beck, Kristin and Biener, Juergen and H{\"a}ffner, Hartmut},
	date = {2025/09/01},
	doi = {10.1038/s41586-025-09474-1},
	id = {Xu2025},
	isbn = {1476-4687},
	journal = {Nature},
	number = {8080},
	pages = {362--368},
	title = {{3D-printed micro ion trap technology for quantum information applications}},
	url = {https://doi.org/10.1038/s41586-025-09474-1},
	volume = {645},
	year = {2025}}

@article{BARQUEST201718,
	author = {B.R. Barquest and G. Bollen and P.F. Mantica and K. Minamisono and R. Ringle and S. Schwarz and C.S. Sumithrarachchi},
	doi = {https://doi.org/10.1016/j.nima.2017.05.036},
	issn = {0168-9002},
	journal = {Nucl. Instrum. Methods Phys. Res. A},
	pages = {18-28},
	title = {{RFQ beam cooler and buncher for collinear laser spectroscopy of rare isotopes}},
	url = {https://www.sciencedirect.com/science/article/pii/S0168900217305892},
	volume = {866},
	year = {2017}}

@article{ZHANG2017297,
	author = {Zai-Yue Zhang and Guang-Zhou Yuan and Yang He and Jie Qian and Shu-Guang Zhang and Ru-Jiao Yao and Chuan-Fan Ding and Xiao-Xu Li},
	doi = {https://doi.org/10.1016/S1872-2040(17)60995-2},
	issn = {1872-2040},
	journal = {Chin. J. Anal. Chem.},
	number = {2},
	pages = {297-302},
	title = {{Research of Unidirectional Ion Ejection in Printed-Circuit-Board Ion Trap}},
	url = {https://www.sciencedirect.com/science/article/pii/S1872204017609952},
	volume = {45},
	year = {2017}}

@article{PhysRevA.75.015401,
  title = {{Loading and characterization of a printed-circuit-board atomic ion trap}},
  author = {Brown, Kenneth R. and Clark, Robert J. and Labaziewicz, Jaroslaw and Richerme, Philip and Leibrandt, David R. and Chuang, Isaac L.},
  journal = {Phys. Rev. A},
  volume = {75},
  issue = {1},
  pages = {015401},
  numpages = {4},
  year = {2007},
  month = {Jan},
  publisher = {American Physical Society},
  doi = {10.1103/PhysRevA.75.015401},
  url = {https://link.aps.org/doi/10.1103/PhysRevA.75.015401}
}

@article{10.1063/1.3665647,
	author = {Narayanan, S. and Daniilidis, N. and M{\"o}ller, S. A. and Clark, R. and Ziesel, F. and Singer, K. and Schmidt-Kaler, F. and H{\"a}ffner, H.},
	doi = {10.1063/1.3665647},
	journal = {J. Appl. Phys.},
	month = {12},
	number = {11},
	pages = {114909},
	title = {{Electric field compensation and sensing with a single ion in a planar trap}},
	url = {https://doi.org/10.1063/1.3665647},
	volume = {110},
	year = {2011}}

@book{coombs2016printed,
  author    = {Coombs, Jr., Clyde F. and Holden, Happy},
  title     = {{Printed Circuits Handbook}},
  edition   = {7th},
  publisher = {McGraw-Hill Education},
  year      = {2016},
  isbn      = {978-0071833950}
}

@article{VALVERDE2020330,
	author = {A.A. Valverde and M. Brodeur and J.A. Clark and D. Lascar and G. Savard},
	doi = {https://doi.org/10.1016/j.nimb.2019.04.070},
	issn = {0168-583X},
	journal = {Nucl. Instrum. Methods Phys. Res. B},
	pages = {330-333},
	title = {{A cooler-buncher for the N=126 factory at Argonne National Laboratory}},
	url = {https://www.sciencedirect.com/science/article/pii/S0168583X19302447},
	volume = {463},
	year = {2020}}

@incollection{DEHMELT196853,
	author = {H.G. Dehmelt},
	doi = {https://doi.org/10.1016/S0065-2199(08)60170-0},
	editor = {D.R. Bates and Immanuel Estermann},
	issn = {0065-2199},
	pages = {53-72},
	publisher = {Academic Press},
	series = {Advances in Atomic and Molecular Physics},
	title = {{Radiofrequency Spectroscopy of Stored Ions I: Storage}},
	url = {https://www.sciencedirect.com/science/article/pii/S0065219908601700},
	volume = {3},
	year = {1968}}

@article{PhysRevLett.96.253003,
  title = {Microfabricated Surface-Electrode Ion Trap for Scalable Quantum Information Processing},
  author = {Seidelin, S. and Chiaverini, J. and Reichle, R. and Bollinger, J. J. and Leibfried, D. and Britton, J. and Wesenberg, J. H. and Blakestad, R. B. and Epstein, R. J. and Hume, D. B. and Itano, W. M. and Jost, J. D. and Langer, C. and Ozeri, R. and Shiga, N. and Wineland, D. J.},
  journal = {Phys. Rev. Lett.},
  volume = {96},
  issue = {25},
  pages = {253003},
  numpages = {4},
  year = {2006},
  month = {Jun},
  publisher = {American Physical Society},
  doi = {10.1103/PhysRevLett.96.253003},
  url = {https://link.aps.org/doi/10.1103/PhysRevLett.96.253003}
}

@article{atoms11110139,
	pages = {139},
	author = {Wang, Jun-Ying and Huang, Wen-Xue and Tian, Yu-Lin and Wang, Yong-Sheng and Wang, Yue and Zhang, Wan-Li and Huang, Yuan-Jun and Gan, Zai-Guo and Xu, Hu-Shan},
	doi = {10.3390/atoms11110139},
	issn = {2218-2004},
	journal = {Atoms},
	number = {11},
	title = {{A Radio-Frequency Ion Trap System for the Multi-Reflection Time-of-Flight Mass Spectrometer at SHANS and Its Offline Commissioning}},
	url = {https://www.mdpi.com/2218-2004/11/11/139},
	volume = {11},
	year = {2023}}

@article{Jiang_PCB_massfilter,
	author = {Jiang, Dan and Jiang, Gong-Yu and Li, Xiao-Xu and Xu, Fu-xing and Wang, Liang and Ding, Li and Ding, Chuan-Fan},
	date = {2013/06/18},
	doi = {10.1021/ac400864k},
	isbn = {0003-2700},
	journal = {Analytical Chemistry},
	journal1 = {Analytical Chemistry},
	journal2 = {Anal. Chem.},
	month = {06},
	number = {12},
	pages = {6041--6046},
	publisher = {American Chemical Society},
	title = {{Printed Circuit Board Ion Trap Mass Analyzer: Its Structure and Performance}},
	type = {doi: 10.1021/ac400864k},
	url = {https://doi.org/10.1021/ac400864k},
	volume = {85},
	year = {2013},
	year1 = {2013}}

@article{Schlottmann2019,
	author = {Schlottmann, Florian and Allers, Maria and Kirk, Ansgar T. and Bohnhorst, Alexander and Zimmermann, Stefan},
	date = {2019/09/01},
	doi = {10.1007/s13361-019-02241-3},
	id = {Schlottmann2019},
	isbn = {1879-1123},
	journal = {J. Am. Soc. Mass Spectrom.},
	number = {9},
	pages = {1813--1823},
	title = {{A Simple Printed Circuit Board--Based Ion Funnel for Focusing Low m/z Ratio Ions with High Kinetic Energies at Elevated Pressure}},
	url = {https://doi.org/10.1007/s13361-019-02241-3},
	volume = {30},
	year = {2019}}

@article{GINS201924,
	author = {W. Gins and R.D. Harding and M. Baranowski and M.L. Bissell and R.F. {Garcia Ruiz} and M. Kowalska and G. Neyens and S. Pallada and N. Severijns and Ph. Velten and F. Wienholtz and Z.Y. Xu and X.F. Yang and D. Zakoucky},
	doi = {https://doi.org/10.1016/j.nima.2019.01.082},
	issn = {0168-9002},
	journal = {Nucl. Instrum. Methods Phys. Res. A},
	pages = {24-32},
	title = {{A new beamline for laser spin-polarization at ISOLDE}},
	url = {https://www.sciencedirect.com/science/article/pii/S0168900219301536},
	volume = {925},
	year = {2019}}

@article{prestage1989new,
  title={{New ion trap for frequency standard applications}},
  author={Prestage, JD and Dick, Go J and Maleki, L},
  journal={J. Appl. Phys.},
  volume={66},
  number={3},
  pages={1013--1017},
  year={1989},
  publisher={American Institute of Physics},
  doi = {10.1063/1.343486},
  url = {https://doi.org/10.1063/1.343486}
}

@article{PhysRevResearch.4.033229,
  title = {{Doppler and sympathetic cooling for the investigation of short-lived radioactive ions}},
  author = {Sels, S. and Maier, F. M. and Au, M. and Fischer, P. and Kanitz, C. and Lagaki, V. and Lechner, S. and Leistenschneider, E. and Leimbach, D. and Lykiardopoulou, E. M. and Kwiatkowski, A. A. and Manovitz, T. and Vila Gracia, Y. N. and Neyens, G. and Plattner, P. and Rothe, S. and Schweikhard, L. and Vilen, M. and Wolf, R. N. and Malbrunot-Ettenauer, S.},
  journal = {Phys. Rev. Res.},
  volume = {4},
  issue = {3},
  pages = {033229},
  numpages = {19},
  year = {2022},
  month = {Sep},
  publisher = {American Physical Society},
  doi = {10.1103/PhysRevResearch.4.033229},
  url = {https://link.aps.org/doi/10.1103/PhysRevResearch.4.033229}
}

@article{VIRTANEN2025170186,
	author = {V.A. Virtanen and T. Eronen and A. Kankainen and O. Beliuskina and P. Campbell and R. Delaplanche and Z. Ge and R.P. {de Groote} and M. Hukkanen and A. Jaries and R. Kronholm and I.D. Moore and A. Raggio and A. de Roubin and J. Ruotsalainen and M. Schuh},
	doi = {https://doi.org/10.1016/j.nima.2024.170186},
	issn = {0168-9002},
	journal = {Nucl. Instrum. Methods Phys. Res. A},
	pages = {170186},
	title = {{Miniaturised cooler-buncher for reduction of longitudinal emittance at IGISOL}},
	url = {https://www.sciencedirect.com/science/article/pii/S0168900224011124},
	volume = {1072},
	year = {2025}}

@article{RANJAN201587,
title = {Design, construction and cooling system performance of a prototype cryogenic stopping cell for the Super-FRS at FAIR},
journal = {Nucl. Instrum. Methods Phys. Res. A},
volume = {770},
pages = {87-97},
year = {2015},
issn = {0168-9002},
doi = {https://doi.org/10.1016/j.nima.2014.09.075},
url = {https://www.sciencedirect.com/science/article/pii/S0168900214011073},
author = {M. Ranjan and P. Dendooven and S. Purushothaman and T. Dickel and M.P. Reiter and S. Ayet and E. Haettner and I.D. Moore and N. Kalantar-Nayestanaki and H. Geissel and W.R. Plaß and D. Schäfer and C. Scheidenberger and F. Schreuder and H. Timersma and J. {Van de Walle} and H. Weick}
}

@article{BARQUEST2016207,
	author = {B.R. Barquest and J.C. Bale and J. Dilling and G. Gwinner and R. Kanungo and R. Kr{\"u}cken and M.R. Pearson},
	doi = {https://doi.org/10.1016/j.nimb.2016.02.035},
	issn = {0168-583X},
	journal = {Nucl. Instrum. Methods Phys. Res. B},
	pages = {207-210},
	title = {Development of a new RFQ beam cooler and buncher for the CANREB project at TRIUMF},
	url = {https://www.sciencedirect.com/science/article/pii/S0168583X16001658},
	volume = {376},
	year = {2016}}

@article{HERFURTH2001254,
	author = {F Herfurth and J Dilling and A Kellerbauer and G Bollen and S Henry and H.-J Kluge and E Lamour and D Lunney and R.B Moore and C Scheidenberger and S Schwarz and G Sikler and J Szerypo},
	doi = {https://doi.org/10.1016/S0168-9002(01)00168-1},
	issn = {0168-9002},
	journal = {Nucl. Instrum. Methods Phys. Res. A},
	number = {2},
	pages = {254-275},
	title = {{A linear radiofrequency ion trap for accumulation, bunching, and emittance improvement of radioactive ion beams}},
	url = {https://www.sciencedirect.com/science/article/pii/S0168900201001681},
	volume = {469},
	year = {2001}}

@article{PhysRevResearch.3.043041,
  title = {{Hypersonic nozzle for laser-spectroscopy studies at 17 K characterized by resonance-ionization-spectroscopy-based flow mapping}},
  author = {Ferrer, R. and Verlinde, M. and Verstraelen, E. and Claessens, A. and Huyse, M. and Kraemer, S. and Kudryavtsev, Yu. and Romans, J. and Van den Bergh, P. and Van Duppen, P. and Zadvornaya, A. and Chazot, O. and Grossir, G. and Kalikmanov, V. I. and Nabuurs, M. and Reynaerts, D.},
  journal = {Phys. Rev. Res.},
  volume = {3},
  issue = {4},
  pages = {043041},
  numpages = {18},
  year = {2021},
  month = {Oct},
  publisher = {American Physical Society},
  doi = {10.1103/PhysRevResearch.3.043041},
  url = {https://link.aps.org/doi/10.1103/PhysRevResearch.3.043041}
}

@article{YU2024169371,
title = {{A laser ablation carbon cluster ion source for the FRS Ion Catcher}},
journal = {Nucl. Instrum. Methods Phys. Res. A},
volume = {1064},
pages = {169371},
year = {2024},
issn = {0168-9002},
doi = {https://doi.org/10.1016/j.nima.2024.169371},
url = {https://www.sciencedirect.com/science/article/pii/S0168900224002973},
author = {Jiajun Yu and Christine Hornung and Timo Dickel and Wolfgang R. Plaß and Daler Amanbayev and Julian Bergmann and Zhuang Ge and Florian Greiner and Hans Geissel and Lizzy Gröf and Gabriella Kripko-Koncz and Meetika Narang and Ann-Kathrin Rink and Christoph Scheidenberger and Jianwei Zhao}
}

@article{Plass_2015,
doi = {10.1088/0031-8949/2015/T166/014069},
year = {2015},
month = {nov},
publisher = {IOP Publishing},
volume = {2015},
number = {T166},
pages = {014069},
author = {Plaß, Wolfgang R and Dickel, Timo and Andres, Samuel Ayet San and Ebert, Jens and Greiner, Florian and Hornung, Christine and Jesch, Christian and Lang, Johannes and Lippert, Wayne and Majoros, Tamas and Short, Devin and Geissel, Hans and Haettner, Emma and Reiter, Moritz P and Rink, Ann-Kathrin and Scheidenberger, Christoph and Yavor, Mikhail I},
title = {{High-performance multiple-reflection time-of-flight mass spectrometers for research with exotic nuclei and for analytical mass spectrometry}},
journal = {Physica Scripta}
}

@article{MORIN2026166027,
title = {{An experimental setup for the study of gas-cell processes for the S3-Low Energy Branch}},
journal = {Nucl. Instrum. Methods Phys. Res. B},
volume = {573},
pages = {166027},
year = {2026},
issn = {0168-583X},
doi = {https://doi.org/10.1016/j.nimb.2026.166027},
url = {https://www.sciencedirect.com/science/article/pii/S0168583X26000285},
author = {E. Morin and W. Dong and V. Manea and A. Claessens and S. Damoy and R. Ferrer and S. Franchoo and S. Geldhof and T. Hourat and Yu. Kudryavtsev and N. Lecesne and R. Leroy and D. Lunney and V. Marchand and E. {Minaya Ramirez} and S. Raeder and S. Roset and Ch. Vandamme and P. {Van den Bergh} and P. {Van Duppen}}
}

@article{DICKEL2015172,
	author = {T. Dickel and W.R. Pla{\ss} and A. Becker and U. Czok and H. Geissel and E. Haettner and C. Jesch and W. Kinsel and M. Petrick and C. Scheidenberger and A. Simon and M.I. Yavor},
	doi = {https://doi.org/10.1016/j.nima.2014.12.094},
	issn = {0168-9002},
	journal = {Nucl. Instrum. Methods Phys. Res. A},
	pages = {172-188},
	title = {{A high-performance multiple-reflection time-of-flight mass spectrometer and isobar separator for the research with exotic nuclei}},
	url = {https://www.sciencedirect.com/science/article/pii/S0168900214015629},
	volume = {777},
	year = {2015}}

@article{JARIES2025170273,
	author = {A. Jaries and J. Ruotsalainen and R. Kronholm and T. Eronen and A. Kankainen},
	doi = {https://doi.org/10.1016/j.nima.2025.170273},
	issn = {0168-9002},
	journal = {Nucl. Instrum. Methods Phys. Res. A},
	pages = {170273},
	title = {{HIBISCUS: A new ion beam radio-frequency quadrupole cooler-buncher for high-precision experiments with exotic radioactive ions}},
	url = {https://www.sciencedirect.com/science/article/pii/S0168900225000749},
	volume = {1073},
	year = {2025}}

@article{102711_094939,
	author = {Dilling, Jens and Blaum, Klaus and Brodeur, Maxime and Eliseev, Sergey},
	doi = {https://doi.org/10.1146/annurev-nucl-102711-094939},
	issn = {1545-4134},
	journal = {Annu. Rev. Nucl. Part. Sci.},
	number = {Volume 68, 2018},
	pages = {45-74},
	publisher = {Annual Reviews},
	title = {{Penning-Trap Mass Measurements in Atomic and Nuclear Physics}},
	type = {Journal Article},
	url = {https://www.annualreviews.org/content/journals/10.1146/annurev-nucl-102711-094939},
	volume = {68},
	year = {2018}}

@article{CAMPBELL2016127,
	author = {P. Campbell and I.D. Moore and M.R. Pearson},
	doi = {https://doi.org/10.1016/j.ppnp.2015.09.003},
	issn = {0146-6410},
	journal = {Progress in Particle and Nuclear Physics},
	pages = {127-180},
	title = {{Laser spectroscopy for nuclear structure physics}},
	url = {https://www.sciencedirect.com/science/article/pii/S0146641015000915},
	volume = {86},
	year = {2016}}

@article{GERBAUX2023167631,
	author = {M. Gerbaux and P. Ascher and A. Husson and A. {de Roubin} and P. Alfaurt and M. Aouadi and B. Blank and L. Daudin and S. {El Abbeir} and M. Flayol and H. Gu{\'e}rin and S. Gr{\'e}vy and M. Hukkanen and B. Lachacinski and D. Lunney and S. Perard and B. Thomas},
	doi = {https://doi.org/10.1016/j.nima.2022.167631},
	issn = {0168-9002},
	journal = {Nucl. Instrum. Methods Phys. Res. A},
	pages = {167631},
	title = {{The General Purpose Ion Buncher: A radiofrequency quadrupole cooler-buncher for DESIR at SPIRAL2}},
	url = {https://www.sciencedirect.com/science/article/pii/S0168900222009238},
	volume = {1046},
	year = {2023}}

@article{SCHWARZ2016131,
	author = {S. Schwarz and G. Bollen and R. Ringle and J. Savory and P. Schury},
	doi = {https://doi.org/10.1016/j.nima.2016.01.078},
	issn = {0168-9002},
	journal = {Nucl. Instrum. Methods Phys. Res. A},
	pages = {131-141},
	title = {The LEBIT ion cooler and buncher},
	url = {https://www.sciencedirect.com/science/article/pii/S0168900216001194},
	volume = {816},
	year = {2016}}

@article{risp_2017,
	author = {Boussaid, Ramzi and Park, Young-Ho and Kondrashev, Sergey},
	date = {2017/12/01},
	doi = {10.3938/jkps.71.848},
	id = {Boussaid2017},
	isbn = {1976-8524},
	journal = {J. Korean Phys. Soc.},
	number = {11},
	pages = {848--854},
	title = {{Technical design of RISP RFQ Cooler buncher}},
	url = {https://doi.org/10.3938/jkps.71.848},
	volume = {71},
	year = {2017}}

@article{BRUNNER201232,
	author = {T. Brunner and M.J. Smith and M. Brodeur and S. Ettenauer and A.T. Gallant and V.V. Simon and A. Chaudhuri and A. Lapierre and E. Man{\'e} and R. Ringle and M.C. Simon and J.A. Vaz and P. Delheij and M. Good and M.R. Pearson and J. Dilling},
	doi = {https://doi.org/10.1016/j.nima.2012.02.004},
	issn = {0168-9002},
	journal = {Nucl. Instrum. Methods Phys. Res. A},
	pages = {32-43},
	title = {TITAN's digital RFQ ion beam cooler and buncher, operation and performance},
	url = {https://www.sciencedirect.com/science/article/pii/S0168900212001398},
	volume = {676},
	year = {2012}}

@article{PODADERAALISEDA2004647,
	author = {I. {Podadera Aliseda} and T. Fritioff and T. Giles and A. Jokinen and M. Lindroos and F. Wenander},
	doi = {https://doi.org/10.1016/j.nuclphysa.2004.09.043},
	issn = {0375-9474},
	journal = {Nucl. Phys. A},
	pages = {647-650},
	title = {{Design of a second generation RFQ Ion Cooler and Buncher (RFQCB) for ISOLDE}},
	url = {https://www.sciencedirect.com/science/article/pii/S0375947404009868},
	volume = {746},
	year = {2004}}

@book{yavor2009optics,
  author    = {Yavor, Mikhail},
  title     = {Optics of Charged Particle Analyzers},
  series    = {Advances in Imaging and Electron Physics},
  volume    = {157},
  publisher = {Academic Press},
  address   = {San Diego},
  year      = {2009},
  isbn      = {978-0-12-374768-6}
}

@article{BARLOW200119,
title = {{Determination of analytic potentials from finite element computations}},
journal = {Int. J. Mass Spectrom.},
volume = {207},
number = {1},
pages = {19-29},
year = {2001},
issn = {1387-3806},
doi = {https://doi.org/10.1016/S1387-3806(00)00452-8},
url = {https://www.sciencedirect.com/science/article/pii/S1387380600004528},
author = {S.E. Barlow and A.E. Taylor and K. Swanson}
}

@article{DAHL20003,
title = {{SIMION for the personal computer in reflection}},
journal = {Int. J. Mass Spectrom.},
volume = {200},
number = {1},
pages = {3-25},
year = {2000},
issn = {1387-3806},
doi = {https://doi.org/10.1016/S1387-3806(00)00305-5},
url = {https://www.sciencedirect.com/science/article/pii/S1387380600003055},
author = {David A Dahl}
}

@article{10.1063/1.1715212,
    author = {Wiley, W. C. and McLaren, I. H.},
    title = {Time‐of‐Flight Mass Spectrometer with Improved Resolution},
    journal = {{Review of Scientific Instruments}},
    volume = {26},
    number = {12},
    pages = {1150-1157},
    year = {1955},
    month = {12},
    issn = {0034-6748},
    doi = {10.1063/1.1715212},
    url = {https://doi.org/10.1063/1.1715212}
}

@article{REITER2021165823,
title = {{Commissioning and performance of TITAN’s Multiple-Reflection Time-of-Flight Mass-Spectrometer and isobar separator}},
journal = {{Nucl. Instrum. Methods Phys. Res. A}},
volume = {1018},
pages = {165823},
year = {2021},
issn = {0168-9002},
doi = {https://doi.org/10.1016/j.nima.2021.165823},
url = {https://www.sciencedirect.com/science/article/pii/S0168900221008081},
author = {M.P. Reiter and S. {Ayet San Andrés} and J. Bergmann and T. Dickel and J. Dilling and A. Jacobs and A.A. Kwiatkowski and W.R. Plaß and C. Scheidenberger and D. Short and C. Will and C. Babcock and E. Dunling and A. Finlay and C. Hornung and C. Jesch and R. Klawitter and B. Kootte and D. Lascar and E. Leistenschneider and T. Murböck and S.F. Paul and M. Yavor},
}

@article{10.1063/1.1712366,
    author = {Franklin, J. L. and Hierl, Peter M. and Whan, David A.},
    title = {{Measurement of the Translational Energy of Ions with a Time‐of‐Flight Mass Spectrometer}},
    journal = {{J. Chem. Phys.}},
    volume = {47},
    number = {9},
    pages = {3148-3153},
    year = {1967},
    month = {11},
    issn = {0021-9606},
    doi = {10.1063/1.1712366},
    url = {https://doi.org/10.1063/1.1712366}
}

@phdthesis{brown2024critical,
  title={Critical quantum phase transitions, investigated through first-time mass measurements of exotic ytterbium},
  author={Brown, Callum Lewis},
  year={2024},
  school={The University of Edinburgh},
  URL = {http://dx.doi.org/10.7488/era/5483}
}

@article{https://doi.org/10.1002/rcm.735,
author = {Douglas, D. J. and Konenkov, N. V.},
title = {Influence of the 6th and 10th spatial harmonics on the peak shape of a quadrupole mass filter with round rods},
journal = {{Rapid Commun. Mass Spectrom.}},
volume = {16},
number = {15},
pages = {1425-1431},
doi = {https://doi.org/10.1002/rcm.735},
url = {https://analyticalsciencejournals.onlinelibrary.wiley.com/doi/abs/10.1002/rcm.735},
year = {2002}
}

@article{l1cn-28kv,
  title = {{Electric field distortions in surface ion traps with integrated nanophotonics}},
  author = {Du, Guochun and Jordan, Elena and Mehlst\"aubler, Tanja E.},
  journal = {{Phys. Rev. Appl.}},
  volume = {25},
  issue = {6},
  pages = {064001},
  numpages = {16},
  year = {2026},
  month = {Jun},
  publisher = {American Physical Society},
  doi = {10.1103/l1cn-28kv},
  url = {https://link.aps.org/doi/10.1103/l1cn-28kv}
}

@article{10.1119/5.0243389,
    author = {Thomas, Robert E. and Wolfram, Cole E. and Warren, Noah B. and Fouch, Isaac J. and Blinov, Boris B. and Parsons, Maxwell F.},
    title = {{An accessible planar charged particle trap for experiential learning in quantum technologies}},
    journal = {{Am. J. Phys.}},
    volume = {93},
    number = {7},
    pages = {581-588},
    year = {2025},
    month = {07},
    issn = {0002-9505},
    doi = {10.1119/5.0243389},
    url = {https://doi.org/10.1119/5.0243389}
}

@article{PhysRevA.72.013405,
  title = {{Proposal for a planar Penning ion trap}},
  author = {Castrej\'on-Pita, J. R. and Thompson, R. C.},
  journal = {{Phys. Rev. A}},
  volume = {72},
  issue = {1},
  pages = {013405},
  numpages = {7},
  year = {2005},
  month = {Jul},
  publisher = {American Physical Society},
  doi = {10.1103/PhysRevA.72.013405},
  url = {https://link.aps.org/doi/10.1103/PhysRevA.72.013405}
}

@article{LECHNER2024169471,
title = {{Simulations of a cryogenic, buffer-gas filled Paul trap for low-emittance ion bunches}},
journal = {Nucl. Instrum. Methods Phys. Res. A},
volume = {1065},
pages = {169471},
year = {2024},
issn = {0168-9002},
doi = {https://doi.org/10.1016/j.nima.2024.169471},
url = {https://www.sciencedirect.com/science/article/pii/S0168900224003978},
author = {S. Lechner and S. Sels and I. Belosevic and F. Buchinger and P. Fischer and C. Kanitz and V. Lagaki and F.M. Maier and P. Plattner and L. Schweikhard and M. Vilen and S. Malbrunot-Ettenauer}
}

@article{Varentsov_2025,
publisher = {IOP Publishing},
volume = {20},
number = {08},
pages = {P08031},
author = {Varentsov, Victor},
title = {{Towards a new and compact gas-dynamic cooler-buncher for the FAIR Laspec and MATS experiments}},
journal = {J.~Instrum.},
doi = {10.1088/1748-0221/20/08/P08031},
url = {https://doi.org/10.1088/1748-0221/20/08/P08031},
year = {2025},
month = {aug},
}

@article{Isolde_laser09,
	author = {Man{\'e}, E. and Billowes, J. and Blaum, K. and Campbell, P. and Cheal, B. and Delahaye, P. and Flanagan, K. T. and Forest, D. H. and Franberg, H. and Geppert, C. and Giles, T. and Jokinen, A. and Kowalska, M. and Neugart, R. and Neyens, G. and N{\"o}rtersh{\"a}user, W. and Podadera, I. and Tungate, G. and Vingerhoets, P. and Yordanov, D. T.},
	date = {2009/12/01},
	doi = {10.1140/epja/i2009-10828-0},
	id = {Man{\'e}2009},
	isbn = {1434-601X},
	journal = {Eur. Phys. J. A},
	number = {3},
	pages = {503--507},
	title = {{An ion cooler-buncher for high-sensitivity collinear laser spectroscopy at ISOLDE}},
	url = {https://doi.org/10.1140/epja/i2009-10828-0},
	volume = {42},
	year = {2009}}

@article{isoltrap08,
	author = {Mukherjee, M. and Beck, D. and Blaum, K. and Bollen, G. and Dilling, J. and George, S. and Herfurth, F. and Herlert, A. and Kellerbauer, A. and Kluge, H. -J. and Schwarz, S. and Schweikhard, L. and Yazidjian, C.},
	date = {2008/01/01},
	doi = {10.1140/epja/i2007-10528-9},
	id = {Mukherjee2008},
	isbn = {1434-601X},
	journal = {Eur. Phys. J. A},
	number = {1},
	pages = {1--29},
	title = {{ISOLTRAP: An on-line Penning trap for mass spectrometry on short-lived nuclides}},
	url = {https://doi.org/10.1140/epja/i2007-10528-9},
	volume = {35},
	year = {2008}}

@article{PLA2013134,
	author = {Wolfgang R. Pla{\ss} and Timo Dickel and Christoph Scheidenberger},
	doi = {https://doi.org/10.1016/j.ijms.2013.06.005},
	issn = {1387-3806},
	journal = {Int. J. Mass Spectrom.},
	note = {100 years of Mass Spectrometry},
	pages = {134-144},
	title = {{Multiple-reflection time-of-flight mass spectrometry}},
	url = {https://www.sciencedirect.com/science/article/pii/S138738061300239X},
	volume = {349-350},
	year = {2013}}

@article{Neugart_2017,
	author = {Neugart, R and Billowes, J and Bissell, M L and Blaum, K and Cheal, B and Flanagan, K T and Neyens, G and N{\"o}rtersh{\"a}user, W and Yordanov, D T},
	doi = {10.1088/1361-6471/aa6642},
	journal = {J. Phys. G: Nucl. Part. Phys.},
	month = {apr},
	number = {6},
	pages = {064002},
	publisher = {IOP Publishing},
	title = {{Collinear laser spectroscopy at ISOLDE: new methods and highlights}},
	url = {https://doi.org/10.1088/1361-6471/aa6642},
	volume = {44},
	year = {2017}}

@article{BLAUM20061,
	author = {Klaus Blaum},
	doi = {https://doi.org/10.1016/j.physrep.2005.10.011},
	issn = {0370-1573},
	journal = {Phys. Rep.},
	number = {1},
	pages = {1-78},
	title = {{High-accuracy mass spectrometry with stored ions}},
	url = {https://www.sciencedirect.com/science/article/pii/S0370157305004643},
	volume = {425},
	year = {2006}}

@article{ALLERS201932,
title = {{Printed circuit board based segmented quadrupole ion guide}},
journal = {Int. J. Mass Spectrom.},
volume = {443},
pages = {32-40},
year = {2019},
issn = {1387-3806},
doi = {https://doi.org/10.1016/j.ijms.2019.05.018},
url = {https://www.sciencedirect.com/science/article/pii/S1387380619301472},
author = {Maria Allers and Florian Schlottmann and Manuel Eckermann and Stefan Zimmermann}
}

@PHDTHESIS{Dickel:200291,
      author       = {Dickel, Timo},
      title        = {{D}esign and commissioning of an
                      {U}ltra-{H}igh-{R}esolution {T}ime-of-{F}light based isobar
                      separator and mass spectrometer [05.11.2010]},
      school       = {Justus-Liebig-Universität Gießen},
      address      = {Darmstadt},
      publisher    = {GSI},
      reportid     = {GSI-2016-03524, GSI Diss 2010-16},
      pages        = {130 S.},
      year         = {2011},
      cin          = {BUD / FRS},
      cid          = {I:(DE-Ds200)BUD-20051214OR030 /
                      I:(DE-Ds200)FRS-20110310OR124},
      pnm          = {899 - ohne Topic (POF3-899)},
      pid          = {G:(DE-HGF)POF3-899},
      experiment   = {EXP:(DE-Ds200)Altdaten-20200803},
      typ          = {PUB:(DE-HGF)11},
      url          = {https://repository.gsi.de/record/200291},
}

\end{document}